\documentclass[lettersize,doublecolumn]{IEEEtran}
\IEEEoverridecommandlockouts
\usepackage{lineno}
\usepackage{hyperref}
\usepackage{cite}
\usepackage{amsmath,amssymb,amsfonts,cases} 
\usepackage{amsmath}
\usepackage{amsthm}
\usepackage{subfig}

\usepackage{setspace}
\usepackage{graphicx}
\usepackage{textcomp}
\usepackage{xcolor}
\usepackage{graphicx}
\usepackage{float}
\usepackage{amsmath,mleftright}
\usepackage{amsfonts,amssymb}
\usepackage{mathrsfs}
\usepackage{mathtools}
\usepackage{algorithm}
\usepackage{algorithmic}
\usepackage{bm}
\usepackage{multirow}
\usepackage{array}
\usepackage{amssymb}
\usepackage{amsmath}
\usepackage{cite}
\usepackage{url}
\usepackage{xcolor}
\usepackage{cite,graphicx,amsmath,amssymb}
\usepackage{fancyhdr}
\usepackage{mdwmath}
\usepackage{mdwtab}
\usepackage{caption}
\usepackage{amsthm}
\usepackage{setspace}
\usepackage{bm}
\usepackage{mathtools}
\usepackage{dsfont}
\usepackage{bbm}
\usepackage{framed}
\allowdisplaybreaks[4]
\newtheorem{remark}{Remark}
\newtheorem{theorem}{Theorem}

\newtheorem{lemma}{Lemma}

\newtheorem{corollary}{Corollary}

\makeatletter
\newcommand{\biggg}{\bBigg@{3}}
\newcommand{\Biggg}{\bBigg@{3.5}}
\makeatother
\makeatletter
\renewcommand{\maketag@@@}[1]{\hbox{\m@th\normalsize\normalfont#1}}%
\makeatother
\def\BibTeX{{\rm B\kern-.05em{\sc i\kern-.025em b}\kern-.08em
    T\kern-.1667em\lower.7ex\hbox{E}\kern-.125emX}}
    \expandafter\def\expandafter\normalsize\expandafter{%
    \normalsize%
    \setlength\abovedisplayskip{4pt}%
    \setlength\belowdisplayskip{4pt}%
    \setlength\abovedisplayshortskip{2pt}%
    \setlength\belowdisplayshortskip{2pt}%
}
\newcounter{problem}
\newcounter{save@equation}
\newcounter{save@problem}
\makeatletter
\newenvironment{problem}
{\setcounter{problem}{\value{save@problem}}%
  \setcounter{save@equation}{\value{equation}}%
  \let\c@equation\c@problem
  \subequations
}
{\endsubequations
  \setcounter{save@problem}{\value{equation}}%
  \setcounter{equation}{\value{save@equation}}%
}

\begin{document}
\title{Exploiting Segmented Waveguide-Enabled Pinching-Antenna Systems (SWANs) for Uplink Tri-Hybrid Beamforming}
 \author{Hao Jiang, Chongjun Ouyang, Zhaolin Wang, Yuanwei Liu,~\IEEEmembership{Fellow,~IEEE}, Arumugam Nallanathan,~\IEEEmembership{Fellow,~IEEE}, Zhiguo Ding,~\IEEEmembership{Fellow,~IEEE}, and 
 Robert Schober,~\IEEEmembership{Fellow,~IEEE}
 \thanks{Hao Jiang, Chongjun Ouyang, and Arumugam Nallanathan are with the School of Electronic Engineering and Computer Science, Queen Mary University of London, London, E1 4NS, U.K. (email: \{hao.jiang, c.ouyang, a.nallanathan\}@qmul.ac.uk).}
 \thanks{Zhaolin Wang and Yuanwei Liu are with the Department of Electrical and Computer Engineering, The University of Hong Kong, Hong Kong (email: \{zhaolin.wang, yuanwei\}@hku.hk).}
\thanks{Z. Ding is with the School of Electrical and Electronic Engineering (EEE), Nanyang Technological University, Singapore 639798 (e-mail: zhiguo.ding@ntu.edu.sg).}
 \thanks{R. Schober is with the Institute for Digital Communications, Friedrich Alexander-University Erlangen-Nurnberg (FAU), Germany (e-mail: robert.schober@fau.de). }
 }

\maketitle
\begin{abstract}
A segmented waveguide-enabled pinching-antenna system (SWAN)-based tri-hybrid beamforming architecture is proposed for uplink multi-user MIMO communications, which jointly optimizes digital, analog, and pinching beamforming.
Both fully-connected (FC) and partially-connected (PC) structures between RF chains and segment feed points are considered.
For the FC architecture, tri-hybrid beamforming is optimized using the weighted minimum mean-square error (WMMSE) and zero-forcing (ZF) approaches. 
Specifically, the digital, analog, and pinching beamforming components are optimized via a closed-form solution, Riemannian manifold optimization, and a Gauss-Seidel search, respectively.
For the PC architecture, an interleaved topology tailored to the SWAN receiver is proposed, in which segments assigned to each RF chain (sub-array) are interleaved with those from other sub-arrays. 
Based on this structure, a WMMSE-based tri-hybrid design is developed, in which the Riemannian-manifold update used for the FC structure is replaced by element-wise phase calibration to exploit sparsity in analog beamforming.
To gain insight into the performance of the proposed system, the rate-scaling laws with respect to the number of segments are derived for both the FC and PC structures.
Our results demonstrate that: i)~SWAN with the proposed tri-hybrid beamforming consistently outperforms conventional hybrid beamforming and conventional pinching-antenna systems with pinching beamforming for both the FC and PC structures; and ii)~the PC structure can strike a good balance between sum rate and energy consumption when the number of segments is large; and iii) the achievable rate does not necessarily increase with the number of segments. 
\end{abstract}

\section{Introduction}
Over the past decades, multiple-input multiple-output (MIMO) technology has played a pivotal role in telecommunications, providing fast and ubiquitous global connectivity \cite{lu2014overviewmimo, alkhateeb2014mimo}.
Advances in MIMO have delivered greater array, multiplexing, and diversity gains, satisfying the growing demands for higher spectral efficiency, lower latency, and enhanced reliability \cite{ning2026precoding, shafi20175g}.

Research on MIMO can be categorized into three primary technical trends: i) Larger antenna apertures, ii) denser antenna placements, and iii) reconfigurable antenna architectures.
In particular, larger antenna apertures integrate hundreds or thousands of antenna elements with a half-wavelength separation, leading to very large physical aperture sizes, as exemplified by massive MIMO, extremely large-scale antenna arrays (ELAAs), and gigantic MIMO \cite{liu2025near, björnson2026antenna}.
In contrast, denser antenna placements increase the aperture density by removing the half-wavelength constraint between adjacent elements, which enables the formation of holographic or continuous surfaces over fixed aperture sizes \cite{gong2024holographic}.

Although promising, the first two trends continue to rely on passive adaptation to wireless channels, resulting in increased hardware costs and implementation complexity.
As a remedy, reconfigurable antenna architectures capable of manipulating wireless channels have attracted significant attention recently \cite{liu2025reconfigurable}.
These reconfigurable antennas enable task-specific adaptation of wireless propagation, either by antenna repositioning or port selection.
Integrated with the existing hybrid beamforming architecture, reconfigurable antennas can be added as a third layer, serving as a dedicated mechanism for manipulating the electromagnetic (EM) properties of wireless channels \cite{heath2025trihybrid}. 
Consequently, a tri-hybrid beamforming architecture comprising digital, analog, and reconfigurable EM beamforming is realized, offering an additional degree of freedom (DoF) for beamforming design.
By leveraging this new DoF, low-cost, high-energy-efficiency MIMO deployments can be provided for next-generation wireless networks \cite{castellanos2026embarcing}. 

\subsection{Prior Works}
To realize a tri-hybrid beamforming architecture, different types of reconfigurable antennas can be exploited, including movable antennas (MAs) \cite{zhu2024movable}, fluid antenna systems (FASs) \cite{new2025fulid_tutorial}, and dynamic meta-surfaces (DMAs) \cite{shlezinger2021dynamic}.
However, the majority of reconfigurable antennas can only support small-scale channel manipulations, as their reconfigurability relies on wavelength-scale operations, such as port repositioning, polarization adjustments, and frequency selection.
To enable large-scale channel manipulations, NTT DOCOMO has proposed pinching-antenna systems (PASS) \cite{Fukuda2022Pinching}, consisting of waveguides for long-range signal transmission and pinching antennas (PAs) for signal radiation.
Based on these components, PASS deliver intended signals to users via a combination of in-waveguide low-attenuation propagation and flexible free-space propagation \cite{ding2025flexible}.
Unlike existing reconfigurable antenna prototypes, PAs can be repositioned over scales of tens, even hundreds of meters, enabling large-scale channel reconfigurability.

In addition to the conventional beamforming architecture, PASS further allow for antenna position optimization, referred to as pinching beamforming, thereby forming a tri-hybrid beamforming architecture.
As a promising technology for 6G and beyond, PASS has attracted growing worldwide interest from both industry and academia.
In particular, the authors of \cite{wang2025modeling} proposed a physics-based model of in-waveguide propagation, which laid the foundation for PASS-based channel modeling.
On top of this model, a power minimization problem for downlink tri-hybrid beamforming was investigated for both continuous and discrete PA-position setups, which represent the most prevalent methods for implementing pinching beamforming.
Moreover, the authors of \cite{bereyhi2025mimo} considered both uplink and downlink tri-hybrid beamforming for sum-rate maximization, demonstrating the superiority of PASS over conventional MIMO and even massive MIMO.
Building on the above, the authors in \cite{sun2026multiuser} devised an element-wise optimization framework for tri-hybrid beamforming, enabling low-complexity joint baseband and pinching beamforming.
Furthermore, the authors of \cite{xu2025pinching} refined the in-waveguide channel model of PASS and revealed that the proportional power radiated at PAs can be used as a new optimization DoF.
Furthermore, regarding waveguides as a new form of resource block, the authors in \cite{zhao2026pinching} presented several transmission structures for pinching beamforming.
Based on this newly developed power radiation model, the optimal tri-hybrid beamforming solution was identified using a branch-and-bound (BnB) method, demonstrating the real potential of PASS over conventional MIMO.
To overcome the computational complexity induced by pinching beamforming, machine learning (ML)-based solutions, such as graph neural networks (GNNs) and transformers \cite{guo2025graph}, were developed to offload optimization to the offline training stage, enabling low-complexity online implementations.

Moreover, the flexibility of PASS-enabled tri-hybrid beamforming can be leveraged for specific applications, including physical-layer security (PLS), integrated sensing and communications (ISAC), and simultaneous wireless information and power transfer (SWIPT), see also \cite{liu2025pinchingantenna}.

\subsection{Motivations and Contributions}
Despite promising early results, existing work primarily focuses on the fully digital implementation of the tri-hybrid beamforming architecture, thus raising concerns regarding the resulting prohibitive hardware cost.

To mitigate this issue, hybrid digital-analog beamforming can be leveraged to reduce the number of RF chains and has been regarded as a practical enabler for massive MIMO \cite{molisch2017hybrid}.
To harness the benefits of hybrid beamforming, the authors of \cite{zhao2025trihybrid} considered a sum-rate maximization problem for a downlink PASS-based tri-hybrid beamforming architecture.
Further, the analytical benefits of PASS-based tri-hybrid beamforming were characterized in \cite{cheng2025performance}.
However, extending this work to the uplink case can be problematic due to inter-antenna radiation (IAR), which occurs when signals received by one PA are re-radiated by the other PAs on the same waveguide.
Moreover, this IAR effect can break the duality between the uplink and downlink channels, thereby compromising channel estimation (CE). 
As a remedy, segmented waveguide-enabled pinching-antenna systems (SWAN) can address these issues by replacing the conventional long-range waveguide with multiple short segmented waveguides, each with dedicated feed points \cite{ouyang2025uplink, jiang2025segment}.
With the reduced segment length, IAR can be mitigated by deploying a single PA per segment.
Moreover, short, segmented waveguides can mitigate in-waveguide loss and facilitate maintainability, making the SWAN architecture a promising candidate for large-scale deployment of PASS.

To leverage the advantages of the SWAN-based tri-hybrid beamforming architecture, we propose an optimization framework for uplink tri-hybrid beamforming design and provide a performance analysis to further characterize its performance limits.
The main contributions of this work are summarized as follows:
\begin{itemize}
    \item We investigate an uplink multi-user MIMO system where the receiver is equipped with a SWAN-based tri-hybrid beamforming architecture. Both fully-connected (FC) and partially-connected (PC) structures between RF chains and segment feed points are considered.
    
    \item For the FC structure, we develop weighted mean square-error (WMMSE)- and zero-forcing (ZF)-based tri-hybrid beamforming strategies for uplink sum-rate maximization. 
    A block coordinate descent (BCD) framework is used to unify: i) Closed-form digital beamforming, ii) Riemannian-manifold optimization for analog beamforming, and iii) Gauss–Seidel updates for pinching beamforming.
    
    \item For the PC structure, we propose an interleaved topology tailored to the SWAN receiver, where segments assigned to each RF chain (i.e., sub-array) are interleaved with those from other sub-arrays.
    Building on this, we further develop a WMMSE-based tri-hybrid beamforming method suitable for the PC structure. 
    In particular, within the same BCD framework, we replace the Riemannian-manifold update with element-wise phase calibration to leverage the sparse analog structure and reduce complexity.
    
    \item We further analyze the rate-scaling law of both the FC and PC structures as a function of the number of segments to characterize their performance limits.
    The results unveil that the theoretical rate does not necessarily increase with the number of segments.
    
    \item Extensive simulations show that: i) SWAN-based tri-hybrid beamforming outperforms massive MIMO with conventional hybrid beamforming and PASS with pinching beamforming; and ii) the proposed methods can effectively design tri-hybrid beamforming for both structures.
\end{itemize}

\subsection{Organization and Notations}
The remainder of this paper is organized as follows:
Section \ref{sect:system_model} characterizes the uplink channels and formulates the sum-rate maximization problem.
Building on the above, Section \ref{sect:mmse_fully} and Section \ref{sect:mmse_partial} propose tri-hybrid beamforming algorithms for the FC and PC structures, respectively.
Section \ref{sect:preformance_analysis} sheds light on the performance limits for both structures.
Numerical results are presented in Section \ref{sect:results}. Finally, the paper is concluded in Section \ref{sect:conclusion}.

\emph{Notations:} Scalars, vectors, and matrices are denoted by
lower-case, bold-face lower-case, and bold-face upper-case letters, respectively.
$\mathbb{C}^{M\times N}$ and $\mathbb{R}^{M \times N}$ denote the space of $M\times N$ complex- and real-valued matrices. 
$(\cdot)^{\textsf{T}}$, $(\cdot)^{\textsf{H}}$, $(\cdot)^{\mathscr{*}}$, and $(\cdot)^{-1}$ denote the transpose, conjugate-transpose, conjugate, and inverse operations, respectively.
$[\cdot]_{i,:}$, $[\cdot]_{:, j}$, and $[\cdot]_{i,j}$ denote the operation to extract the $i$-th row, the $j$-th column, and the $(i,j)$-th entry of a matrix, respectively.
$\odot$ and $\nabla$ denote the Hadamard product and gradient operator, respectively.
$\Re\{\cdot\}$ and $\Im\{\cdot\}$ denote the real and imaginary parts of a complex-valued variable, respectively.
$\mathrm{j}=\sqrt{-1}$ and $\mathrm{e}$ denote the imaginary unit and the Euler number, respectively.
$\mathbb{E}\{\cdot\}$ denotes the expectation operator.
$\| \cdot \|$, $|\cdot|$, and $(\cdot)^{|\cdot|}$ denote the $L_2$-norm, the absolute value, and the element-wise absolute value, respectively.
The operator $\angle (\cdot)$ represents the extraction of the phase of a complex number.

\section{System Model} \label{sect:system_model}
\begin{figure}[t!]
	\centering
	\includegraphics[width=0.9\linewidth]{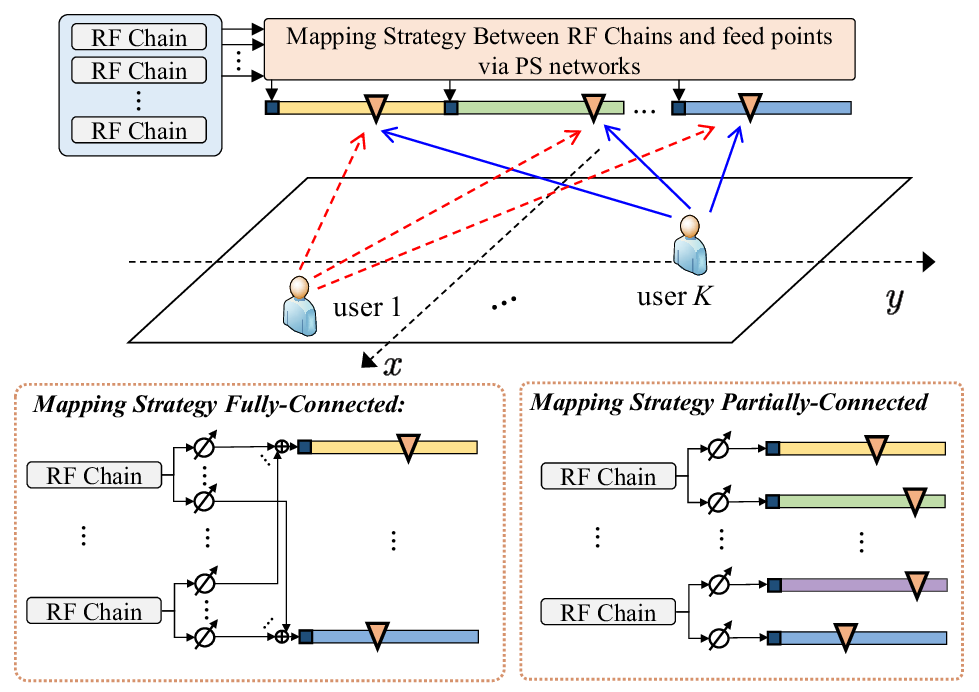}
	\caption{Illustration for multi-user uplink tri-hybrid beamforming architecture proposed for SWAN.}
	\label{fig:system_model}
	\vspace{-2em}
\end{figure}

The system model for SWAN-based tri-hybrid beamforming is shown in Fig. \ref{fig:system_model}, where $K$ single-antenna users transmit signals to a SWAN receiver composed of $M$ segments.
The SWAN service area is defined by $D_x \times D_y$, where $D_x$ and $D_y$ represent the side lengths along the $x$- and $y$-axes, respectively.
Each segment contains one PA, and its length is $L = D_x/M$.
The segments are connected sequentially at a height of $H$ and are oriented parallel to the $x$-axis.
The $x$-coordinate of the feed point on the $m$-th segment is denoted by $x_m^{\mathrm{FD}}$, with all feed points positioned at the left endpoints of the segments.
The geometric location of the $k$-th user is given by $\mathbf{r}_k = [r_{x,k}, r_{y,k}, 0]^{\textsf{T}}$, while the coordinate of the $m$-th PA is $\mathbf{p}_m = [x_m, 0, H]^{\textsf{T}}$.
Since all PAs share common $y$ and $z$-coordinates, i.e., $y_m = 0$ and $z_m = H$ for all $m$, the $x$-coordinate vector $\mathbf{x} = [x_1, ..., x_M]^{\textsf{T}} \in \mathbb{R}^{M \times 1}$ is defined to compactly represent the positions of the PAs.
From a beamforming architecture perspective, two connection strategies between feed points and RF chains are considered: FC and PC.
In the FC structure, each RF chain is connected to all feed points via phase shifters (PSs), whereas in the PC structure, each RF chain is connected to only a subset of feed points.
The number of RF chains and phase shifters are denoted by $N_{\mathrm{RF}}$ and $N_{\mathrm{PS}}$, respectively.

\subsection{System Model}
In this sub-section, we model the uplink channel for the SWAN-based tri-hybrid beamforming architecture, including both free-space and in-waveguide propagation.
In particular, the free-space channel vector from the $k$-th user to the PAs on the SWAN is given by
\begin{align}
	\tilde{\mathbf{h}}_k(\mathbf{x})=\left[ \frac{\sqrt{\eta}}{r_{1,k}}\mathrm{e}^{-\mathrm{j}\frac{2\pi}{\lambda _{\mathrm{c}}}r_{1,k}},...,\frac{\sqrt{\eta}}{r_{M,k}}\mathrm{e}^{-\mathrm{j}\frac{2\pi}{\lambda _{\mathrm{c}}}r_{M,k}} \right] ^{\textsf{T}},
\end{align}
where $\eta=\lambda_{\mathrm{c}}^2/(16\pi^2)$ denotes the free-space path gain constant with $\lambda_{\mathrm{c}}$ being the wavelength in free space, and $r_{m,k}=\| \mathbf{p}_m - \mathbf{r}_k\|$ denotes the Euclidean distance between the $k$-th user and the $m$-th PA.
After being received by the PAs, the signal will undergo in-waveguide propagation to the feed points.
Letting $\Delta _m\triangleq |x_{m}^{\mathrm{FD}}-x_{m}^{}|$ be the propagation distance within the $m$-th segment, the in-waveguide channel is given by  
\begin{align}
   \mathbf{g}(\mathbf{x})=\left[ 10^{-\frac{\kappa \Delta _1}{20}}\mathrm{e}^{-\mathrm{j}\frac{2\pi}{\lambda _{\mathrm{g}}}\Delta _1},...,10^{-\frac{\kappa \Delta _M}{20}}\mathrm{e}^{-\mathrm{j}\frac{2\pi}{\lambda _{\mathrm{g}}}\Delta _M} \right] ^{\textsf{T}},
\end{align}
where $\lambda_{\mathrm{g}}=\lambda_{\mathrm{c}}/n_{\mathrm{eff}}$ denotes the in-waveguide wavelength, $\kappa$ denotes the in-waveguide loss measured in $\mathrm{dB/m}$, and $n_{\mathrm{eff}}$ denotes the effective refractive index of the dielectric waveguide.
Hence, the overall uplink channel for the $k$-th user is defined as follows:
\begin{align}
    {\mathbf{h}}_k(\mathbf{x}) = \tilde{\mathbf{h}}_k(\mathbf{x}) \odot  \mathbf{g}(\mathbf{x}) \in \mathbb{C}^{M \times 1}. \label{eq:overall_channel}
\end{align}
The channel reconfiguration is introduced by designing the PA positions in \eqref{eq:overall_channel}.

Based on the proposed channel model, the received signal at the SWAN receiver can be written as follows:
\begin{align}
	\mathbf{y}=\sum\nolimits_{k=1}^K{\mathbf{h}_k\left( \mathbf{x} \right)}s_k+\mathbf{n},
\end{align}
where $s_k$ denotes the symbol transmitted by the $k$-th user satisfying $\mathbb{E}\{|s_k|^2\} = P$, $P$ denotes the uplink transmit power of the users, and $\mathbf{n} \sim \mathcal{CN}(\boldsymbol{0}_M, \sigma^2 \mathbf{I}_{M})$ denotes the additive Gaussian noise with a power of $\sigma^2$.
More compactly, the received signal can be written as follows:
\begin{align}
	\mathbf{y}=\mathbf{H}(\mathbf{x})\mathbf{s}+\mathbf{n}, \label{eq:received_signal}
\end{align}
where $\mathbf{H}(\mathbf{x})\triangleq[\mathbf{h}_1(\mathbf{x}), ..., \mathbf{h}_K(\mathbf{x})]\in\mathbb{C}^{M \times K}$ is  the uplink channel matrix, and $\mathbf{s}=[s_1, s_2, ..., s_K]^{\textsf{T}}$ is the transmitted symbol vector.
The received signal in \eqref{eq:received_signal} is processed by analog and digital beamforming.
To characterize the uplink performance, we use the achievable uplink sum rate as a metric in this work \cite{bereyhi2025mimo, hou2025on}.
Here, we define the RF and digital beamforming matrices as $\mathbf{G}_{\rm RF}\in \mathbb{C}^{M \times N_{\mathrm{RF}}}$ and $\mathbf{G}_{\mathrm{BB}}\in\mathbb{C}^{N_{\mathrm{RF}}\times K}$, respectively.
The demodulated symbol of the $k$-th user can be expressed as follows:
\begin{align}
	&\hat{s}_k=\mathbf{g}_{k}^{\textsf{H}}\mathbf{G}_{\rm RF}^{\textsf{H}}\mathbf{y}=\underset{\mathrm{D}-\mathrm{BF}}{\underbrace{\mathbf{g}_{k}^{\textsf{H}}}}~\cdot~\underset{\mathrm{A}-\mathrm{BF}} {\underbrace{\mathbf{G}_{\rm RF}^{\textsf{H}}}} ~\cdot~\underset{\mathrm{P}-\mathrm{BF}}{\underbrace{\mathbf{H}(\mathbf{x})}}\mathbf{s}+\mathbf{g}_{k}^{\textsf{H}}\mathbf{G}_{\rm RF}^{\textsf{H}}\mathbf{n}, \label{eq:signal_model}
\end{align}
where $\mathbf{g}_k = [\mathbf{G}_{\rm BB}]_{:, k}$ denotes the digital beamforming vector for the $k$-th user, and ``D-BF", ``A-BF", and ``P-BF" stand for digital beamforming, analog beamforming, and pinching beamforming, respectively.

\subsection{Problem Formulation}
According to the signal model in \eqref{eq:signal_model}, the received signal-to-interference-plus-noise ratio (SINR) of the $k$-th user can be expressed as follows:
\begin{align}
	&\vspace{-1em} \gamma _k\left( \mathbf{g}_{k}^{},\mathbf{G}_{\rm RF}^{},\mathbf{x} \right) \notag \\
	&=\frac{\left| \mathbf{g}_{k}^{\textsf{H}}\mathbf{G}_{\rm RF}^{\textsf{H}}\mathbf{h}_{k}^{}\left( \mathbf{x} \right) \right|^2}{\sum\nolimits_{i=1,i\ne k}^K{\left| \mathbf{g}_{k}^{\textsf{H}}\mathbf{G}_{\rm RF}^{\textsf{H}}\mathbf{h}_{i}^{}\left( \mathbf{x} \right) \right|^2+\frac{\sigma ^2}{P}\left\| \mathbf{g}_{k}^{\textsf{H}}\mathbf{G}_{\rm RF}^{\textsf{H}} \right\| ^2}}. \label{eq:sinr}
\end{align}
Thus, jointly considering all $K$ users, the sum rate can be compactly written as follows:
\begin{align}
	R\left( \mathbf{G}_{\rm BB}, \mathbf{G}_{\rm RF}, \mathbf{x} \right) = \sum\nolimits_{k=1}^K{\log_2 \left( 1+\gamma _k\left( \mathbf{g}_{k}^{},\mathbf{G}_{\rm RF}^{},\mathbf{x} \right) \right)}. \label{eq:sum_rate}
\end{align}
In this work, we consider two structures, i.e., PC and FC. 
In the FC structure, the output of each RF chain is connected to all feed points of the segments through PSs, as shown in the lower left part of Fig. \ref{fig:system_model}.
In the PC structure, the output of each RF chain is connected to a subset of the feed points of the segments via PSs, as depicted in the lower right part of Fig. \ref{fig:system_model}.
Compared to the FC structure, the analog beamforming matrix for the PC structure is sparse, reducing both the number of optimization variables and the number of required PSs.
Consequently, the PC structure enables a low-complexity, low-energy-consumption implementation of hybrid beamforming \cite{yu2016alternating}, though it offers fewer DoFs for optimization due to the sparse analog beamforming matrix.
Based on the above discussion, we formulate the SWAN-based uplink tri-hybrid beamforming problem.
Let $\mathbf{W}_{\rm BB}$ and $\mathbf{F}_{\rm BB}$ denote the digital beamforming matrices for the FC and PC structures, respectively, and let $\mathbf{W}_{\rm RF}$ and $\mathbf{F}_{\rm RF}$ denote the corresponding analog beamforming matrices.
Hence, for the FC structure, we define $(\mathbf{G}_{\rm RF}, \mathbf{G}_{\rm BB}) = (\mathbf{W}_{\rm RF}, \mathbf{W}_{\rm BB})$, whereas, for the PC structure, we define $(\mathbf{G}_{\rm RF}, \mathbf{G}_{\rm BB}) = (\mathbf{F}_{\rm RF}, \mathbf{F}_{\rm BB})$.
With these definitions, the uplink sum-rate maximization problem is formulated as follows: 
\begin{problem}\label{pd:uplink_sum_rate} 
	\begin{align}
		\underset{\mathbf{x}, \mathbf{G}_{\rm RF}, \mathbf{G}_{\rm BB} }{\max} 
		\quad & R\!\left( \mathbf{G}_{\rm RF}, \mathbf{G}_{\rm BB}, \mathbf{x} \right) \\
		\text{s.t.}\quad &  \left|[\mathbf{G}_{\rm RF}]_{i,j}\right| = 1,\ \forall i,j, \label{1tst:3} \\
& \left( \mathbf{G}_{\mathrm{RF}},\mathbf{G}_{\mathrm{BB}} \right) =\begin{cases}
	\left( \mathbf{W}_{\mathrm{RF}},\mathbf{W}_{\mathrm{BB}} \right) ,&		\mathrm{for}~\mathrm{FC},\\
	\left( \mathbf{F}_{\mathrm{RF}},\mathbf{F}_{\mathrm{BB}} \right) ,&		\mathrm{for}~\mathrm{PC},\\
\end{cases} \label{1tst:1} \\
		& \mathbf{x} \in \mathcal{F}. \label{1tst:5}
	\end{align}
\end{problem}
Constraint \eqref{1tst:3} represents the unit-modulus constraint on the analog beamformer; constraint \eqref{1tst:1} denotes the specific realization of the RF beamforming matrix;
and constraint \eqref{1tst:5} represents the position constraint on each PA specified as follows:
\begin{align}
	\mathcal{F} \triangleq \left\{ \mathbf{x}\left| \begin{array}{c}
		(m-1)L\le [\mathbf{x}]_m\le mL,\forall m\in \mathcal{M}\\
		\left| [\mathbf{x}]_m-[\mathbf{x}]_{m^{\prime}} \right|\ge \Delta _{\min},\forall m\ne m^{\prime}\\
	\end{array} \right. \right\},
\end{align}
where $\mathcal{M}$ is defined as $\mathcal{M}\triangleq\{1,2,..., M\}$, and $\Delta_{\min}$ denotes the minimal inter-spacing between two PAs to mitigate the mutual coupling effect \cite{ouyang2025array}.

\section{Solutions for SWAN-Based Tri-Hybrid Beamforming for the FC structure}\label{sect:mmse_fully}
In this section, we first present the weighted minimum mean-square error (WMMSE)-based solution to problem \eqref{pd:uplink_sum_rate}.
Given the widespread use of zero-forcing (ZF) in practical multi-user MIMO systems, a ZF-based solution to the original problem is also presented to enhance the applicability of the proposed SWAN-based uplink tri-hybrid beamforming architecture.

\subsection{WMMSE-Based Solution to \eqref{pd:uplink_sum_rate}} \label{subsect:mmse_fully_w_bb}
In this method, we use a BCD method to optimize the digital beamforming matrix $\mathbf{W}_{\mathrm{BB}}$, analog beamforming matrix $\mathbf{W}_{\mathrm{RF}}$, and PA position vector $\mathbf{x}$ sequentially.
Assuming that $\mathbb{E} \left\{ |s_k|^2 \right\} =P$, $\mathbb{E} \left\{ \mathbf{n}s^{*} \right\} =\boldsymbol{0}_M$, and $\mathbb{E} \left\{ \mathbf{nn}^{\textsf{H}} \right\} =\sigma ^2\mathbf{I}_M$ are satisfied, the mean squared error (MSE) for the $k$-th user can be derived as follows:
\begin{align}
  &e_k=\mathbb{E} \left\{ \left| \hat{s}_k-s_k \right|^2 \right\} \notag \\
  &=P\left| \mathbf{w}_{k}^{\textsf{H}}\mathbf{W}_{\mathrm{RF}}^{\textsf{H}}\mathbf{h}_k(\mathbf{x}) \right|^2+P\sum\nolimits_{j=1,j\ne k}^K{\left| \mathbf{w}_{k}^{\textsf{H}}\mathbf{W}_{\mathrm{RF}}^{\textsf{H}}\mathbf{h}_j(\mathbf{x}) \right|^2}\notag 
  \\
  &-2P\Re \left\{ \mathbf{w}_{k}^{\textsf{H}}\mathbf{W}_{\mathrm{RF}}^{\textsf{H}}\mathbf{h}_k(\mathbf{x}) \right\} +\sigma ^2\left\| \mathbf{w}_{k}^{\textsf{H}}\mathbf{W}_{\mathrm{RF}}^{\textsf{H}} \right\| ^2+P. \label{eq:mse}
\end{align}
It can be seen that the MSE in \eqref{eq:mse} admits a quadratic form and, therefore, is more tractable for optimization than the original sum-log expression.
Furthermore, as demonstrated in \cite{shi2011an} and \cite{lin2019hybrid}, the original sum-log maximization problem is equivalent to the following formulation:
\begin{problem}\label{pd:uplink_sum_rate_1} 
	\begin{align}
			\underset{\mathcal{V}\cup {\Omega}}{\min}\quad & \sum\nolimits_{k=1}^K \left( { \omega _k e_k }-\log \left( \omega _k \right) \right) \label{obj_func:mse} \\
			\text{s.t.}\quad & \eqref{1tst:3}~\text{and}~\eqref{1tst:5}, \notag
		\end{align}
\end{problem}
where $\mathcal{V} = \{\mathbf{W}_{\rm BB}, \mathbf{W}_{\rm RF}, \mathbf{x}\}$ is defined as the set of optimization variables, and $\Omega \triangleq \{\omega_1, ..., \omega_K\}$ denotes a set of weights for all $K$ users.
According to the proof in \cite{shi2011an}, problem \eqref{pd:uplink_sum_rate_1} is equivalent to \eqref{pd:uplink_sum_rate}.
Based on the tractable form of \eqref{pd:uplink_sum_rate_1}, we adopt the BCD framework to decompose the original problem into sub-problems w.r.t. each optimization variable.

\subsubsection{Digital Beamforming Optimization}
In light of the rationale of BCD, we fix optimization variables $\mathbf{x}$ and $\mathbf{W}_{\rm RF}$ and combine the analog beamforming matrix with the channel vector to form the equivalent channel vector, which is given by $\check{\mathbf{h}}_k\triangleq \mathbf{W}_{\mathrm{RF}}^{\textsf{H}} \mathbf{h}_k\left( \mathbf{x} \right) $ for $\forall k$.
Then, the optimal digital beamforming vector for the $k$-th user under the equivalent channel $\check{\mathbf{h}}_k$ can be obtained from the first-order optimality condition of \eqref{obj_func:mse}, which can be written as follows:
\begin{align}
	\nabla_{\mathbf{w}_k^*} e_k &=\notag \\
	&\sum\nolimits_{j=1}^K{P\check{\mathbf{h}}_j\check{\mathbf{h}}_{j}^{\textsf{H}}\mathbf{w}_{k}^{}}-P\check{\mathbf{h}}_k+\sigma ^2\mathbf{W}_{\mathrm{RF}}^{\textsf{H}}\mathbf{W}_{\mathrm{RF}}^{}\mathbf{w}_{k}^{}=\boldsymbol{0}_{N_{\rm RF}}, \notag
\end{align}
from which the optimal beamforming vector can be derived as follows:
\begin{align}
	\mathbf{w}_{k, \star}^{}=\left( \sum\nolimits_{j=1}^K{\check{\mathbf{h}}_j\check{\mathbf{h}}_{j}^{\textsf{H}}}+(\sigma ^2/P)\mathbf{W}_{\mathrm{RF}}^{\textsf{H}}\mathbf{W}_{\mathrm{RF}}^{} \right) ^{-1}\check{\mathbf{h}}_k. \label{eq:opt_mmse}
\end{align}
More compactly, the optimal RF-domain beamforming matrix can be written as $\mathbf{W}_{\rm BB, \star} = [\mathbf{w}_{1, \star}, ...,  \mathbf{w}_{K, \star}]$.

\subsubsection{Analog Beamforming Optimization} \label{subsect:mmse_fully_w_rf}
With the optimal digital beamforming matrix $\mathbf{W}_{\rm BB, \star}$ at hand, we now consider the optimization of the analog beamforming matrix. 
Specifically, plugging $\mathbf{W}_{\rm BB, \star}$ back into the MSE expression in \eqref{eq:mse}, we have 
\begin{align}
	e_{k, \mathbf{w}_{k, \star}}&=P\mathbf{w}_{k,\star}^{\textsf{H}}\mathbf{W}_{\mathrm{RF}}^{\textsf{H}}\left( \sum\nolimits_{j=1}^K{\mathbf{h}_j\mathbf{h}_{j}^{\textsf{H}}}+(\sigma ^2/P)\mathbf{I}_M \right) \notag \\
	&\times \mathbf{W}_{\mathrm{RF}}^{}\mathbf{w}_{k,\star}^{} -2P\Re \left\{ \mathbf{w}_{k,\star}^{\textsf{H}}\mathbf{W}_{\mathrm{RF}}^{\textsf{H}}\mathbf{h}_k \right\} +P .
\end{align}
Therefore, the sub-problem with respect to $\mathbf{W}_{\rm RF}$ can be formulated in the following form:
\begin{problem}\label{pd:uplink_sum_rate_4} 
	\begin{align}
		\underset{\mathbf{W}_{\rm RF}}{\min}\quad & \sum\nolimits_{k=1}^K { \omega _k e_{k, \mathbf{w}_{k, \star}} } \label{obj_func:mse_Wrf_1} \\
		\text{s.t.}\quad & \eqref{1tst:3}. \notag
	\end{align}
\end{problem}
Furthermore, defining $\mathbf{R}\triangleq \sum\nolimits_{j=1}^K{\mathbf{h}_j\mathbf{h}_{j}^{\textsf{H}}}+(\sigma ^2/P)\mathbf{I}_M$, $\mathbf{C}\triangleq P\sum\nolimits_{k=1}^K{\omega_k\mathbf{w}_{k,\star}^{}\mathbf{w}_{k,\star}^{\textsf{H}}}$, and $\mathbf{B}\triangleq 2P\sum\nolimits_{k=1}^K{\omega_k\mathbf{h}_k\mathbf{w}_{k,\star}^{\textsf{H}}}$, the MSE expression for the $k$-th user can be re-written in a more compact form, i.e., 
\begin{align}
f(\mathbf{W}_{\rm RF}) &\triangleq \sum\nolimits_{k=1}^K { \omega _k e_{k, \mathbf{w}_{k, \star}} }\notag \\
    &=P\sum\nolimits_{k=1}^K{\omega_k\mathrm{tr}\left\{ \mathbf{W}_{\mathrm{RF}}^{\textsf{H}}\mathbf{RW}_{\mathrm{RF}}^{}\mathbf{w}_{k,\star}^{}\mathbf{w}_{k,\star}^{\textsf{H}} \right\}}\notag\\
	&-2P\sum\nolimits_{k=1}^K{\Re \left\{ \mathrm{tr}\left\{ \mathbf{W}_{\mathrm{RF}}^{\textsf{H}}\mathbf{h}_k\mathbf{w}_{k,\star}^{\textsf{H}} \right\} \right\}} + KP \notag \\
	&=P\mathrm{tr}\left\{ \mathbf{W}_{\mathrm{RF}}^{\textsf{H}}\mathbf{RW}_{\mathrm{RF}}^{}\sum\nolimits_{k=1}^K{\mathbf{w}_{k,\star}^{}\mathbf{w}_{k,\star}^{\textsf{H}}} \right\} \notag \\
	&-2P\Re \left\{ \mathrm{tr}\left\{ \mathbf{W}_{\mathrm{RF}}^{\textsf{H}}\sum\nolimits_{k=1}^K{\mathbf{h}_k\mathbf{w}_{k,\star}^{\textsf{H}}} \right\} \right\} + KP\notag \\
	&=\mathrm{tr}\left\{ \mathbf{W}_{\mathrm{RF}}^{\textsf{H}}\mathbf{RW}_{\mathrm{RF}}^{}\mathbf{C} \right\} -\Re \left\{ \mathrm{tr}\left\{ \mathbf{W}_{\mathrm{RF}}^{\textsf{H}}\mathbf{B} \right\} \right\} +KP, 
\end{align}
where $\mathrm{tr}\{\mathbf{AB}\}=\mathrm{tr}\{\mathbf{BA}\}$ is invoked.

The crux of solving problem \eqref{pd:uplink_sum_rate_4} originates from the unit-modulus constraint \eqref{1tst:3}, which is not convex.
To address the unit-modulus constraint, we utilize the Riemannian manifold optimization technique to convert problem \eqref{pd:uplink_sum_rate_4} to an unconstrained form \cite{yu2016alternating, zhao2025trihybrid}.
In particular, the unit-modulus constraint defines $M \times N$ complex circles with unit radius, which is a type of manifold where the inner product is defined as $\left< x_1,x_2 \right> =\Re \left\{ x_{1}^{*}x_2 \right\}$.
Based on this definition, the manifold is characterized $\mathcal{M}_\mathrm{cc}=\{|[\mathbf{W}_{\rm RF}]_{i,j}|=1,~\forall i,j \}$. 
Consequently, performing gradient descent on this manifold inherently satisfies the unit-modulus constraint, as every point on the manifold fulfills this requirement.
Gradient descent on a manifold comprises the following steps: 1) Computing the Riemannian gradient, 2) Updating towards the negative direction of the Riemannian gradient, and 3) Retracting the endpoint back onto the manifold.
Following this order, the first step is to compute the Riemannian gradient induced by the Euclidean gradient.
Ignoring constant terms in \eqref{eq:sum_error}, the Euclidean gradient can be obtained as follows:
\begin{align}
    \nabla_{\mathbf W_{\mathrm{RF}}} f(\mathbf{W}_{\rm RF})= 2\,\mathbf R\,\mathbf W_{\mathrm{RF}}\mathbf C- \mathbf B.
\end{align}
Therefore, letting $t$ be the iteration index for the gradient descent process, the Riemannian gradient can be derived as 
\begin{align}
	&\mathbf{G}^{(t)} = \mathrm{grad}~f(\mathbf{W}_{\rm RF}^{(t)}) = \mathrm{proj}_{\mathbf{W}_{\rm RF}}~\nabla _{\mathbf{W}_{\mathrm{RF}}^{(t)}}f\left( \mathbf{W}_{\mathrm{RF}^{(t)}}^{} \right)\notag \\
	& =\nabla _{\mathbf{W}_{\mathrm{RF}}}f(\mathbf{W}_{\mathrm{RF}}^{(t)})-\Re \{\nabla _{\mathbf{W}_{\mathrm{RF}}^{(t)}}f(\mathbf{W}_{\mathrm{RF}}^{(t)})\odot (\mathbf{W}_{\mathrm{RF}}^{t})^*\}\odot \mathbf{W}_{\mathrm{RF}}^{(t)}, \label{eq:remannian_fc}
\end{align}  
which can be interpreted as a projection from the Euclidean gradient onto the tangent space of the manifold $\mathcal{M}$.
Therefore, to ensure that the updated $\mathbf{W}_{\rm RF}$ is still on the manifold, we define the retraction operation as follows \cite{yu2016alternating}:
\begin{align}
\mathbf{W}_{\mathrm{RF}}^{(t+1)}&=\mathrm{retract}\!\left( \mathbf{W}_{\mathrm{RF}}^{(t)}+\alpha \,\mathbf{D}^{\left( t \right)} \right) 
=\frac{\mathbf{W}_{\mathrm{RF}}^{(t)}+\alpha \,\mathbf{D}^{\left( t \right)}}{( \mathbf{W}_{\mathrm{RF}}^{(t)}+\alpha \,\mathbf{D}^{\left( t \right)} ) ^{\left| \cdot \right|}},
\label{eq:retraction_2}
\end{align}
where $\alpha>0$ is the step size and $\mathbf{D}^{(t)}$ denotes a matrix in the tangent space at point $\mathbf{W}_{\rm RF}^{(t)}$.
Furthermore, to facilitate the conjugate gradient, we define the transport operation that maps two tangent spaces at distinct points on $\mathcal{M}_{\rm cc}$.
According to \cite{yu2016alternating}, this operation is defined as follows:
\begin{align}
    \mathbf{D}_{+}^{\left( t \right)}&=\mathrm{trans}(\mathbf{D}_{}^{\left( t \right)}) \notag \\
    &=\mathbf{D}_{}^{\left( t \right)}-\Re \left\{ \mathbf{D}_{}^{\left( t \right)}\odot \left( \mathbf{W}_{\mathrm{RF}}^{(t+1)} \right) ^* \right\} \odot \mathbf{W}_{\mathrm{RF}}^{(t+1)}, \label{eq:transport}
\end{align}
where $\mathbf{D}_{}^{\left( t \right)}$ denotes a matrix in the tangent space at point $\mathbf{W}_{\rm RF}^{(t)}$, while $\mathbf{D}_{+}^{\left( t \right)}$ denotes a matrix in the tangent space at point $\mathbf{W}_{\rm RF}^{(t+1)}$.
The entire gradient descent method is summarized in \textbf{Algorithm \ref{alg:remannian_manifold}}, which computes the optimized RF-domain beamforming matrix denoted by $\mathbf{W}_{\rm RF, \star}$.
\begin{algorithm}[t!] 
    \small
    \caption{Riemannian Gradient Algorithm}
    \label{alg:remannian_manifold}
    \begin{algorithmic}[1]
        \STATE{Initialize the iteration index $t=0$ and the optimization variables $\mathbf{W}_{\rm BB, \star}$, $\mathbf{x}$, and $\mathbf{W}_{\rm RF}^{(0)}$}
        \STATE{Compute first descent direction $\mathbf{D}^{(0)}=-\mathrm{grad}f(\mathbf{W}_{\rm RF}^{(0)})$\;}
        \REPEAT
            \STATE{Choose step size $\alpha_t$ by Armijo backtracking line search\;}
            \STATE{Identify the next point $\mathbf{W}_{\rm RF}^{(t+1)}$ by retraction \eqref{eq:retraction_2} \;}
            \STATE{Compute the Riemannian gradient $\mathbf{G}^{(t+1)}$ by \eqref{eq:remannian_fc} \;}
            \STATE{Compute the transports $\mathbf{G}^{(t)}_+$ and $\mathbf{D}^{(t)}_+$ of $\mathbf{G}^{(t)}$ and $\mathbf{D}^{(t)}$ from $\mathbf{W}_{\rm RF}^{(t)}$ to $\mathbf{W}_{\rm RF}^{(t+1)}$ by \eqref{eq:transport}\;}
            \STATE{Compute Polak–Ribière parameter by $\beta ^{(t+1)}=\max \left\{ \frac{\Re \left\{ \mathrm{tr}\left\{ \left( \mathbf{G}_{+}^{(t)} \right) ^{\textsf{H}}\left( \mathbf{G}_{+}^{(t)}-\mathbf{G}_{}^{(t)} \right) \right\} \right\}}{\Re \left\{ \left( \mathbf{G}_{}^{(t)} \right) ^{\textsf{H}}\mathbf{G}_{}^{(t)} \right\}},0 \right\} $\;}
            \STATE{Compute the conjugate direction by $\mathbf{D}^{(t+1)} = -\mathbf{G}^{(t+1)} + \beta^{(t+1)}\mathbf{D}^{(t)}_+$ \;}
            \STATE{Set $t = t+1$\;}
        \UNTIL{the fractional decrease of the sum-rate falls below a predefined threshold}
        \RETURN{Optimized analog beamforming matrix $\mathbf{W}_{\rm RF, \star} \triangleq \mathbf{W}_{\rm RF}^{(t)}$}
    \end{algorithmic}
\end{algorithm}

\begin{algorithm}[t!]
    \small
    \caption{Gauss-Seidel Algorithm}
    \label{alg:element_wise}
    \begin{algorithmic}[1]
        \STATE{Initialize the iteration index $t=0$ and the optimization variables $\mathbf{W}_{\rm BB, \star}$, $\mathbf{W}_{\rm RF, \star}$, $\mathbf{x}^{\mathrm{(0)}}$}
        \REPEAT
            \FOR{$m \in  \{1,\dots,M\}$}
            \STATE{Update $x_{m}^{(t)}$ by decreasing the MSE \eqref{obj_func:mse}, while checking the feasibility of $\mathbf{x}^{(t)}\in\mathcal{F}$\;}
            \ENDFOR
            \STATE{Set $t = t+1$\;}
        \UNTIL{the fractional decrease of the MSE falls below a predefined threshold}
         \RETURN{Optimized PA positions $\mathbf{x}_{\star} \triangleq \mathbf{x}_{}^{(t)}$}
    \end{algorithmic}
\end{algorithm}
\subsubsection{Pinching Beamforming Optimization} \label{subsect:mmse_fully_x}
For fixed $\mathbf{W}_{\mathrm{BB}, \star}$ and $\mathbf{W}_{\mathrm{RF}, \star}$, the last optimization step aims at finding the optimal PA positions.
To solve this problem, a Gauss-Seidel approach is employed to overcome the multi-modal optimization objective in PASS \cite{wang2025modeling}.
In particular, the $x$-coordinate region of each segment is divided into a set of discrete candidate positions. 
A one-dimensional search is used to traverse these candidate positions. 
The PA position is updated only if the MSE decreases while satisfying the feasibility constraints. 
This procedure is repeated until the relative MSE between consecutive iterations decreases below a prescribed threshold.
The resulting algorithm is summarized in \textbf{Algorithm \ref{alg:element_wise}}, whose output is denoted by $\mathbf{x}_{\star}$.

\subsubsection{Auxiliary Variable Set Optimization} \label{subsect:mmse_fully_omega}
Inserting $\mathbf{W}_{\rm BB, \star}$, $\mathbf{W}_{\rm RF, \star}$, and $\mathbf{x}_{\star}$ obtained by the aforementioned steps into the expression of $e_k$ in \eqref{eq:mse}, $\forall k$, will yield $e _{k, \star}$, $\forall k$.
Thus, the auxiliary variable $\omega_k$, $\forall k$, can be updated by finding the stationary point of function \eqref{obj_func:mse} with respect to (w.r.t.) $\omega_k$, $\forall k$, which yields the following equation:
\begin{align}
	\frac{\partial}{\partial \omega _k}\sum\nolimits_{k=1}^K{\left( \omega _ke_k-\log \left( \omega _{k} \right) \right)}=e_k-{1}/{\omega _k}=0.
\end{align} 
The above equation provides the update rule for $\omega_k$, which is given by
\begin{align}
	\omega _{k,\star}^{}=1/e_{k,\star}. \label{eq:auxiliary_update}
\end{align}
Therefore, we have $\Omega_{\star}=\{\omega_{k, \star}\}_{k=1}^K$.

Based on the discussion in Sections \ref{subsect:mmse_fully_w_bb}, \ref{subsect:mmse_fully_w_rf}, \ref{subsect:mmse_fully_x}, and \ref{subsect:mmse_fully_omega}, the overall BCD algorithm is summarized in \textbf{Algorithm \ref{alg:overall}}, which can find an local optimal solution to the sum-rate maximization problem.
The overall computational complexity is derived as $\mathcal{O}(I_{\rm iter, 1} (M^3 + I_{\rm iter 2}M +  I_{\rm iter 3}M + K)$, where $ I_{\rm iter 1}$, $ I_{\rm iter 2}$, and $ I_{\rm iter 3}$ denote the numbers of iterations needs for BCD, Riemannian gradient descent, and Gauss-Seidel searching, respectively.

\begin{algorithm}[t!]
    \small
    \caption{FC Tri-Hybrid Beamforming via WMMSE}
    \label{alg:overall}
    \begin{algorithmic}[1]
    \STATE{Initialize the iteration index $t=0$ and the optimization variables $\mathbf{W}_{\rm BB, \star}^{(0)}$, $\mathbf{W}_{\rm RF, \star}^{(0)}$, $\mathbf{x}^{(0)}$}
        \REPEAT
            \STATE{Find $\mathbf{W}_{\rm BB, \star}^{(t)}$ by \eqref{eq:opt_mmse} \;}
            \STATE{Find $\mathbf{W}_{\rm RF, \star}^{(t)}$ by \textbf{Algorithm \ref{alg:remannian_manifold}} \;}
            \STATE{Find $\mathbf{x}_{\star}^{(t)}$ by \textbf{Algorithm \ref{alg:element_wise}}\;}
            \STATE{Find $\Omega_{\star}^{(t)}$ by Eqn. \eqref{eq:auxiliary_update}\;}
            \STATE{Set $t = t+1$\;}
        \UNTIL{the fractional increase of the sum-rate falls below a predefined threshold}
        \RETURN{Optimized beamforming matrices $\mathbf{W}_{\rm BB, \star}^{} \triangleq \mathbf{W}_{\rm BB, \star}^{(t)}$ and $\mathbf{W}_{\rm RF, \star}^{} \triangleq \mathbf{W}_{\rm RF, \star}^{(t)}$, and optimized PA positions $\mathbf{x}_{ \star}^{} \triangleq \mathbf{x}_{\star}^{(t)}$ \;}
    \end{algorithmic}
\end{algorithm}

\subsection{Zero-Forcing-Based Solution to \eqref{pd:uplink_sum_rate}}\label{sub-sect:zf}
For the design of the uplink beamforming matrix, ZF is a widely adopted strategy due to its robustness and low computational complexity, making it an important baseline for evaluation and implementation of the proposed tri-hybrid beamforming architecture \cite{wiesel2008zero,ngo2013energy}.
Therefore, we consider the hybrid beamforming via ZF in this sub-section.
To ensure the feasibility of ZF, we assume that $K \le N_{\rm RF}$.
By imposing the ZF constraint, the inter-user interference can be canceled out.
In this case, the optimization problem can be formulated as follows~\cite{su2022optimal}: 
\begin{problem}\label{pd:uplink_sum_rate_zf} 
	\begin{align}
		\underset{\mathcal{V}}{\max} 
		\quad & \sum\nolimits_{k=1}^K{\log _2\left( 1+\frac{P/\sigma ^2}{\left\| \mathbf{W}_{\mathrm{RF}}^{}\mathbf{w}_{k}^{} \right\| ^2} \right)} \label{1tstzf:1} \\
		\text{s.t.}\quad 
		& \left| [\mathbf{W}_{\mathrm{RF}}]_{i,j} \right|=1,\ \forall i,j  \\
		& \mathbf{W}_{\mathrm{BB}}^{\textsf{H}}\mathbf{W}_{\mathrm{RF}}^{\textsf{H}}\mathbf{H}\left( \mathbf{x} \right) =\mathbf{I}_K \label{st_zf:orthogonal}\\
		& \mathbf{x} \in \mathcal{F}. 
	\end{align}
\end{problem}
It is noted that the additional constraint \eqref{st_zf:orthogonal} enforces the mutual orthogonality between the uplink signals from the users. 
Compared with the complicated fractional form in \eqref{pd:uplink_sum_rate}, the objective function in \eqref{pd:uplink_sum_rate_zf} is simplified due to the constraint \eqref{st_zf:orthogonal}.
Therefore, a BCD framework can be directly applied without the need for re-formulating the optimization problem.
In particular, we sequentially optimize the digital beamforming matrix $\mathbf{W}_{\rm BB}$, analog beamforming matrix $\mathbf{W}_{\rm RF}$, and PA positions $\mathbf{x}$, .

\subsubsection{Digital Beamforming Optimization}
For fixed $\mathbf{W}_{\rm RF}$ and $\mathbf{x}$, the optimization sub-problem for the digital beamforming vector of the $k$-th user can be written as follows:
\begin{problem}\label{pd:uplink_sum_rate_zf_W_bb_user_k} 
  	\begin{align}
  		\underset{\mathbf{w}_{k}}{\min} 
  		\quad & \left\| \mathbf{W}_{\mathrm{RF}}^{}\mathbf{w}_{k}^{} \right\| ^2  \\
  		\text{s.t.}\quad&  \mathbf{w}_{k}^{\textsf{H}}\mathbf{W}_{\mathrm{RF}}^{\textsf{H}}\mathbf{H}\left( \mathbf{x} \right)=\mathbf{e}_k^{\textsf{T}}, \label{st_zf:orthogonal_single_user}
  	\end{align}
\end{problem}
where $\mathbf{e}_k=[\mathbf{I}_K]_{:,k}$ denotes the $k$-th column drawn from an identity matrix of dimension $K \times K$, i.e., $\mathbf{I}_K$.
This sub-problem is convex, since it is a quadratic minimization problem with an affine constraint, and, therefore, can be solved optimally.
According to the Karush-Kuhn-Tucker (KKT) condition, the Lagrangian function can be written as follows:
\begin{align}
	\mathcal{L} \left( \mathbf{w}_{k}^{} \right) &=\mathbf{w}_{k}^{\textsf{H}}\mathbf{W}_{\mathrm{RF}}^{\textsf{H}}\mathbf{W}_{\mathrm{RF}}^{}\mathbf{w}_{k}^{} \notag \\
	&+\boldsymbol{\lambda }_{k}^{\textsf{H}}\left( \check{\mathbf{H}}_{}^{\textsf{H}}\mathbf{w}_{k}^{}-\mathbf{e}_k \right)+\left( \mathbf{w}_{k}^{\textsf{H}}\check{\mathbf{H}}-\mathbf{e}_k \right) \boldsymbol{\lambda }_{k}^{},
\end{align}
where the equivalent channel matrix is defined as $\check{\mathbf{H}}\triangleq \mathbf{W}_{\mathrm{RF}}^{\textsf{H}}\mathbf{H}^{}\left( \mathbf{x} \right) $, and $\boldsymbol{\lambda}_k \in \mathbb{C}^{K \times 1}$ denotes the Lagrangian multipliers.
By exploiting the stationary condition of $\mathcal{L} \left( \mathbf{w}_{k}^{} \right)$, we have $\nabla _{\mathbf{w}_{k}^{*}}\mathcal{L} \left( \mathbf{w}_{k}^{} \right) =\mathbf{W}_{\mathrm{RF}}^{\textsf{H}}\mathbf{W}_{\mathrm{RF}}^{}\mathbf{w}_{k}^{}+\check{\mathbf{H}}\boldsymbol{\lambda }_{k}^{}=\boldsymbol{0}_{\rm N}$.
where $\boldsymbol{0}_{N}\in\mathbb{C}^{N\times 1}$ denotes the all-zero vector.
Hence, the optimal solution of the beamforming vector for the $k$-th user can be expressed as
\begin{align}
	\mathbf{w}_{k}^{}=-\left( \mathbf{W}_{\mathrm{RF}}^{\textsf{H}}\mathbf{W}_{\mathrm{RF}}^{} \right) ^{-1}\check{\mathbf{H}}\boldsymbol{\lambda }_{k}^{}. \label{eq:optimal_w_k_w_lambda}
\end{align}
Then, leveraging constraint \eqref{st_zf:orthogonal_single_user}, the optimal Lagrangian multiplier $\boldsymbol{\lambda}_{k, \star}$ can be computed as 
\begin{align}
	\boldsymbol{\lambda }_{k,\star}^{}=-\left( \check{\mathbf{H}}^{\textsf{H}}\left( \mathbf{W}_{\mathrm{RF}}^{\textsf{H}}\mathbf{W}_{\mathrm{RF}}^{} \right) ^{-1}\check{\mathbf{H}}^{} \right) ^{-1}\mathbf{e}_k. \label{eq:optimal_lambda}
\end{align}
Based on \eqref{eq:optimal_w_k_w_lambda} and \eqref{eq:optimal_lambda}, the optimal solution for the digital beamforming vector of the $k$-th user, i.e., $\mathbf{w}_{k, \star}$ can be derived as follows:
\begin{align}
	\mathbf{w}_{k,\star}^{}=\left( \mathbf{W}_{\mathrm{RF}}^{\textsf{H}}\mathbf{W}_{\mathrm{RF}}^{} \right) ^{-1}\check{\mathbf{H}}\left( \check{\mathbf{H}}^{\textsf{H}}\left( \mathbf{W}_{\mathrm{RF}}^{\textsf{H}}\mathbf{W}_{\mathrm{RF}}^{} \right) ^{-1}\check{\mathbf{H}}^{} \right) ^{-1}\mathbf{e}_k. \notag 
\end{align}
Thus, the optimal digital beamforming matrix is defined as $\mathbf{W}_{\mathrm{BB},\star}=\mathbf{D}^{-1}\check{\mathbf{H}}\left( \check{\mathbf{H}}^{\textsf{H}}\mathbf{D}^{-1}\check{\mathbf{H}}^{} \right) ^{-1}\mathbf{I}_K$, where $\mathbf{D}\triangleq \mathbf{W}_{\rm RF}^{\textsf{H}}\mathbf{W}_{\rm RF}^{}$.

\subsubsection{Analog Beamforming Optimization} 
Given the optimal digital beamforming matrix $\mathbf{W}_{\mathrm{BB},\star}$, we 
optimize $\mathbf{W}_{\rm RF}$ under the unit-modulus constraint.
The optimization problem w.r.t. $\mathbf{W}_{\rm RF}$ can be formulated as follows: 
\begin{problem}\label{pd:uplink_sum_rate_zf_W_rf} 
	\begin{align}
		\underset{\mathbf{W}_{\rm RF}}{\max} 
		~ & f(\mathbf{W}_{\rm RF}) \triangleq \sum_{k=1}^K{\log _2\left( 1+\frac{P/\sigma ^2}{\mathrm{tr}\left\{ \mathbf{E}_k\left( \check{\mathbf{H}}^{\textsf{H}}\mathbf{D}^{-1}\check{\mathbf{H}}^{} \right) ^{-1} \right\}} \right)}  \label{obj:W_rf}\\
		\text{s.t.}~&   \left| [\mathbf{W}_{\mathrm{RF}}]_{i,j} \right|=1,\ \forall i,j,
	\end{align}
\end{problem}
where $\mathbf{E}_k\triangleq \mathbf{e}_k\mathbf{e}_{k}^{\textsf{T}}$ denotes the entry-selection matrix.
To deal with the unit-modulus constraint, we resort to the Riemannian manifold method again.
As detailed in Section \ref{subsect:mmse_fully_w_rf}, the initial step is to compute the Euclidean gradient of the objective function and transform it into the Riemannian manifold.
For the following derivations, the following definitions are useful:
\begin{subequations}
	\begin{align}
		&\mathbf{T}_1 \triangleq \mathbf{D}^{-1}, \quad \mathbf{T}_2=\check{\mathbf{H}}^{\textsf{H}}\mathbf{T}_1\check{\mathbf{H}}^{} ,\quad \mathbf{T}_3 = \mathbf{T}_2^{-1}\\
		&t_k \triangleq \left[ \mathbf{T}_3 \right] _{k,k}, \quad c=P/\sigma^2.
	\end{align}
\end{subequations}
First, $\partial f(\mathbf{W}_{\rm RF}) / \partial r_k$ is given by
\begin{align}
	\frac{\partial f(\mathbf{W}_{\rm RF})}{\partial t_k} = -\frac{c}{\ln(2) t_k(t_k + c)}.
\end{align}
Therefore, by defining $\mathbf{T}_4 \triangleq \mathrm{diag}\{[\partial f / \partial t_1, ..., \partial f / \partial t_K]\}$, $df(\mathbf{W}_{\rm RF})$ can be compactly written as $df(\mathbf{W}_{\rm RF}) = {\mathrm{tr}}\{{\mathbf{T}_4 d(\mathbf{T}_3)}\}$.
According to the chain rule, we further have $d{\mathbf{T}_3} = -\mathbf{T}_3(d\mathbf{T}_2)\mathbf{T}_3$.
Therefore, the overall gradient can be expressed as $df = -\mathrm{tr}\{\mathbf{T}_3\mathbf{T}_4\mathbf{T}_3d\mathbf{T}_2\}$ and can be further expanded as follows:
\begin{align}
	df = -I_1 -I_2 -I_3,
\end{align}
where we use the following definitions $I_1 \triangleq \mathrm{tr}\{\mathbf{T}_3\mathbf{T}_4\mathbf{T}_3\left( d\check{\mathbf{H}}^{\textsf{H}} \right) \mathbf{T}_1\check{\mathbf{H}}^{}\}$, $I_2\triangleq \mathrm{tr}\left\{\mathbf{T}_3\mathbf{T}_4\mathbf{T}_3\check{\mathbf{H}}^{\textsf{H}}\left( d\mathbf{T}_1 \right) \check{\mathbf{H}}^{} \right\}$, and $I_3\triangleq \mathrm{tr}\left\{ \mathbf{T}_3\mathbf{T}_4\mathbf{T}_3\check{\mathbf{H}}^{\textsf{H}}\mathbf{T}_1d\check{\mathbf{H}}^{} \right\}$.
Hence, the exact forms of $I_1$, $I_2$, and $I_3$ are derived. 
In particular, based on the definition of $\check{\mathbf{H}}^{}$, $I_1$ and $I_3$ can be expressed as follows:
\begin{align}
    I_1&=\mathrm{tr}\{\mathbf{T}_1\check{\mathbf{H}}^{}\mathbf{T}_3\mathbf{T}_4\mathbf{T}_3\mathbf{H}^{\textsf{H}}\left( d\mathbf{W}_{\mathrm{RF}} \right) \}, \\
    I_3&=\mathrm{tr}\{\mathbf{HT}_3\mathbf{T}_4\mathbf{T}_3\check{\mathbf{H}}^{\textsf{H}}\mathbf{T}_1\left( d\mathbf{W}_{\mathrm{RF}}^{\textsf{H}} \right) \}.
\end{align}
Accordingly, the following equation holds: $I_1+I_3=2\Re \{\mathrm{tr}\{\mathbf{T}_1\check{\mathbf{H}}^{}\mathbf{T}_3\mathbf{T}_4\mathbf{T}_3\mathbf{H}^{\textsf{H}}\left( d\mathbf{W}_{\mathrm{RF}} \right) \}\}$.
Subsequently, based on the fact that $d\mathbf{T}_1=-\mathbf{T}_1\left( d\mathbf{D} \right) \mathbf{T}_1$ and $d\mathbf{D}=d\mathbf{W}_{\mathrm{RF}}^{\textsf{H}}\mathbf{W}_{\mathrm{RF}}^{}+\mathbf{W}_{\mathrm{RF}}^{\textsf{H}}d\mathbf{W}_{\mathrm{RF}}^{}$, the term $I_2$ can be simplified into the following form: 
\begin{align}
	&I_2=-\mathrm{tr}\left\{ \mathbf{W}_{\mathrm{RF}}^{}\mathbf{T}_1\check{\mathbf{H}}^{}\mathbf{T}_3\mathbf{T}_4\mathbf{T}_3\check{\mathbf{H}}^{\textsf{H}}\mathbf{T}_1d\mathbf{W}_{\mathrm{RF}}^{\textsf{H}} \right\} \notag \\
	&-\mathrm{tr}\left\{ \mathbf{T}_1\check{\mathbf{H}}^{}\mathbf{T}_3\mathbf{T}_4\mathbf{T}_3\check{\mathbf{H}}^{\textsf{H}}\mathbf{T}_1\mathbf{W}_{\mathrm{RF}}^{\textsf{H}}\left( d\mathbf{W}_{\mathrm{RF}}^{} \right) \right\} \notag 
	\\
	&=-2\Re \left\{ \mathrm{tr}\left\{ \mathbf{T}_1\check{\mathbf{H}}^{}\mathbf{T}_3\mathbf{T}_4\mathbf{T}_3\check{\mathbf{H}}^{\textsf{H}}\mathbf{T}_1\mathbf{W}_{\mathrm{RF}}^{\textsf{H}}d\mathbf{W}_{\mathrm{RF}}^{} \right\} \right\}.
\end{align}
Therefore, the Euclidean gradient can be written as follows:
\begin{align}
	&\nabla _{\mathbf{W}_{\mathrm{RF}}}f\left( \mathbf{W}_{\mathrm{RF}}^{} \right) =\notag \\
	&\mathbf{W}_{\mathrm{RF}}^{}\mathbf{T}_1\check{\mathbf{H}}^{}\mathbf{T}_3\mathbf{T}_4\mathbf{T}_3\check{\mathbf{H}}^{\textsf{H}}\mathbf{T}_1-\mathbf{HT}_3\mathbf{T}_4\mathbf{T}_3\check{\mathbf{H}}^{\textsf{H}}\mathbf{T}_1.
\end{align}
Based on the Euclidean gradient, the Riemannian gradient, the retraction operation, and the transport operation can be defined in the same form as in \eqref{eq:remannian_fc}, \eqref{eq:retraction_2}, and \eqref{eq:transport}, respectively.
Given these operations, Riemannian gradient ascent w.r.t. the objective function in \eqref{obj:W_rf} can be executed. The procedure is summarized in \textbf{Algorithm \ref{alg:remannian_manifold}}, where the minus sign in the conjugate directions and the objective function are replaced accordingly.

\subsubsection{Pinching Beamforming Optimization}
Based on the obtained $\mathbf{W}_{\rm BB, \star}$ and $\mathbf{W}_{\rm RF, \star}$, the remaining problem w.r.t. PA position optimizations can be solved by the Gauss-Seidel approach.
The detailed steps have been summarized in \textbf{Algorithm \ref{alg:element_wise}} and are omitted for brevity.

Therefore, the overall algorithm can be summarized as in Algorithm \ref{alg:overall}, except that the update step for the auxiliary variable set is no longer needed.
The computational complexity of the resulting overall BCD algorithm is given by $\mathcal{O}(I_{\rm iter, 1} (M^3 + I_{\rm iter 2}M +  I_{\rm iter 3}M )$, where $ I_{\rm iter 1}$, $ I_{\rm iter 2}$, and $ I_{\rm iter 3}$ denote the number of iterations needed for BCD, Riemannian gradient descent, and Gauss-Seidel searching, respectively.

\section{Solution for SWAN-Based Tri-Hybrid Beamforming for the PC structure} \label{sect:mmse_partial}
Although the FC structure can achieve maximum performance, it still suffers from high computational complexity and reduced energy efficiency resulting from the high-dimensional analog beamforming matrix.
The PC structure can overcome these limitations, as it reduces the number of RF chains through enabling each RF chain to connect to a subset of antennas via PSs.
Consequently, the dimension of analog beamforming is reduced, resulting in fewer phase shifters and a corresponding decrease in energy consumption.
In this section, we first present a PC structure suitable for the proposed SWAN architecture.
Then, we present the tri-hybrid beamforming architecture for the PC structure.

\subsection{Partially-Connected Structure for SWAN}
\begin{figure}
    \centering
    \includegraphics[height=0.7\linewidth]{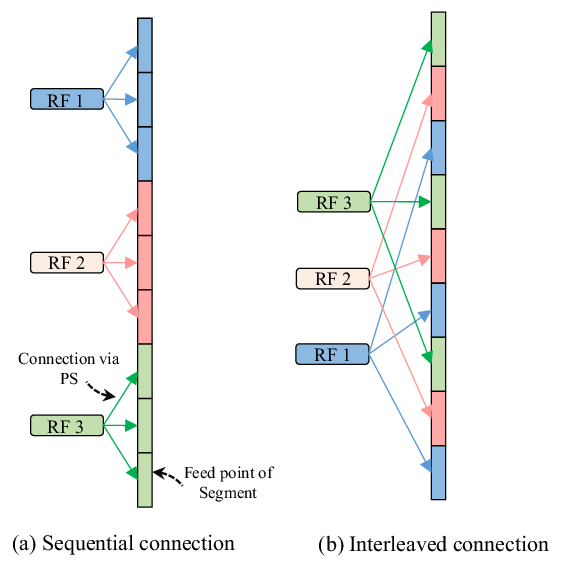}
    \caption{Illustration of the PC structures: Sequential versus interleaved.\vspace{-1em}}
    \label{fig:parially_struct}
\end{figure}

In the conventional fixed-position arrays (FPA)-based PC structure, each RF chain is connected to feed points in a sub-array-by-sub-array manner, as illustrated in Fig. \ref{fig:parially_struct}a.
As shown in this figure, sub-arrays, composed of an RF chain connecting several feed points via PSs, are arranged sequentially.
Therefore, the RF beamforming matrix is given by
\begin{align}
    \mathbf{F}_{\mathrm{RF}}^{\mathrm{FPA}}=\left[ \begin{smallmatrix}
	\mathbf{c}_{1}^{}&		\boldsymbol{0}_{N_{\rm par}}&		\cdots&		\boldsymbol{0}_{N_{\rm par}}\\
	\boldsymbol{0}_{N_{\rm par}}&		\mathbf{c}_{2}^{}&		&		\\
	\vdots&		&		\ddots&		\vdots\\
	\boldsymbol{0}_{N_{\rm par}}&		\boldsymbol{0}_{N_{\rm par}}&		\cdots&		\mathbf{c}_{N_{\mathrm{RF}}}^{}\\
\end{smallmatrix} \right],
\end{align}
where $\mathbf{c}_i = [\mathrm{e}^{\mathrm{j} \phi_{(i-1)N_{\rm par}+1}}, ..., \mathrm{e}^{\mathrm{j} \phi_{iN_{\rm par}}}]^{\textsf{T}} \in \mathbb{C}^{N_{\rm par} \times 1}$ denotes the sub-array connected to the $i$-th RF chain and $\phi_{i} \in (-\pi, \pi]$ denotes the phase of the $i$-th PS.
Here, the number of antennas in one sub-array, i.e., $N_{\rm par}$, is given by $N_{\rm par} = M/N_{\rm RF}$.

This sequential layout is commonly employed in conventional FPA scenarios because the path loss encountered by different sub-arrays is nearly identical.
In contrast, in the SWAN architecture, each segment typically spans several meters.
As a result of the random user distribution and the considerable length of the SWAN architecture, each user can be served by one or multiple sub-arrays, depending on its location.
Hence, the sequential layout will inevitably degrade fairness among users.
To this end, we propose a new PC structure tailored for the SWAN architecture, referred to as the interleaved layout, as shown in Fig. \ref{fig:parially_struct}b.
Specifically, segments connected to different RF chains are arranged in an alternating pattern, which ensures that the sub-array segments are uniformly distributed throughout the SWAN architecture.

To mathematically model this interleaved connection, we define a permutation matrix, denoted by $\mathbf{\Pi}\in\{0,1\}^{M\times M}$, where $M=N_{\mathrm{RF}}N_{\rm par}$ represents the total number of PSs.
Here, we permute the row order of $\mathbf{F}_{\mathrm{RF}}^{\mathrm{FPA}}$ from the sequential index $j=(j^\prime-1)N_{\rm par} + i^\prime$ to the interleaved index $i=(i^\prime-1)N_{\rm par} + j^\prime$, where $i^\prime=1,2,..., N_{\rm par}$ and $j^\prime=1,2,...,N_{\rm RF}$.
Building on the above definitions, the index-pair set can be defined as follows:
\begin{align}
    \mathcal{I} _{\mathrm{pair}}&\triangleq \left\{ \left\{ (i^\prime -1)N_{\mathrm{RF}}+j^\prime , (j^\prime -1)N_{\mathrm{par}}+i^\prime \right\} \right. \notag \\
    &\left. \mid i^\prime =1,2,...,N_{\mathrm{par}}, j^\prime =1,2,...,N_{\mathrm{RF}} \right\}
\end{align}
The $(i,j)$-th entry equal to one in $\mathbf{\Pi}$ represents the operation to permute the $j$-th row of $\mathbf{F}_{\rm RF}^{\rm FPA}$ to the $i$-th row.
In light of this interpretation, the entire permutation matrix is defined as follows:
\begin{align}
    \left[ \mathbf{\Pi } \right] _{i,j}=\begin{cases}
	1,&		\mathrm{if}~i\in \mathcal{I} _{\mathrm{pair}},\\
	0,&		\mathrm{otherwise}.\\
\end{cases}
\end{align}
Consequently, the RF beamforming matrix of the proposed interleaved PC structure is obtained by permuting the feed-point ordering of the conventional block-diagonal matrix, which can be expressed as
\begin{align}
\mathbf{F}_{\mathrm{RF}}^{\mathrm{SWAN}}
=\mathbf{\Pi}\mathbf{F}_{\mathrm{RF}}^{\mathrm{FPA}},
\end{align}
which is a row permutation version of $\mathbf{F}_{\mathrm{RF}}^{\mathrm{FPA}}$.
In what follows, we use $\mathbf{F}_{\mathrm{RF}}$ to denote $\mathbf{F}_{\mathrm{RF}}^{\mathrm{SWAN}}$ for brevity of notation.
Due to the sparse structure of $\mathbf{F}_{\mathrm{RF}}$, the ZF strategy can be sensitive to the condition number of $\mathbf{F}_{\mathrm{RF}}$, thereby degrading performance.
Thus, we present a WMMSE strategy for this scenario, although a ZF approach is also possible by following the same steps as in Section \ref{sub-sect:zf}.

\subsection{WMMSE-Based Solution to \eqref{pd:uplink_sum_rate}}
For the interleaved PC structure, the optimization problem is formulated as follows:
\begin{problem}\label{pd:uplink_sum_rate_partially_connected} 
	\begin{align}
		\underset{\mathcal{V}}{\max} 
		\quad & R\!\left( \mathbf{F}_{\mathrm{BB}}, \mathbf{F}_{\mathrm{RF}}, \mathbf{x} \right) \\
		\text{s.t.}\quad 
		& \left| [\mathbf{F}_{\mathrm{RF}}]_{i,j} \right|=1,\ \forall i,j \\
		& \mathbf{x} \in \mathcal{F},
	\end{align}
\end{problem}
where $\mathbf{F}_{\rm BB}$ denotes the digital beamforming matrix.
To solve this problem, we invoke the WMMSE-based solution elaborated in Section \ref{sect:mmse_fully}.
To make this part self-contained, we briefly present the core steps.
First, for the WMMSE expression, the digital beamforming vector is obtained as
\begin{align}
    \mathbf{f}_{k, \star}^{}=\left( \sum\nolimits_{k=1}^K{P\bar{\mathbf{h}}_k\bar{\mathbf{h}}_k^{\textsf{H}}}+(\sigma ^2/P)\mathbf{F}_{\mathrm{RF}}^{\textsf{H}}\mathbf{F}_{\mathrm{RF}}^{} \right) ^{-1}\bar{\mathbf{h}}_k,
\end{align}
where $\bar{\mathbf{h}}_k\triangleq \mathbf{F}_{\mathrm{RF}}^{\textsf{H}}\mathbf{h}_k\left( \mathbf{x} \right)$.
Consequently, the digital beamforming matrix is constructed by $\mathbf{F}_{\mathrm{BB}, \star}=[\mathbf{f}_{1, \star}^{}, ..., \mathbf{f}_{K, \star}^{}]$.
For the optimization of the analog beamforming matrix, the sparse structure of $\mathbf{F}_{\rm RF}$ prevents the application of the Riemannian optimization approach, as the definition of the manifold is compromised. 
Therefore, to fully leverage the sparse structure and reduce computational complexity, we use an element-wise phase-matching method to determine the PS values in $\mathbf{F}_{\rm RF}$.
More specifically, plugging $\mathbf{F}_{\rm BB, \star}$ back into the MSE expression in \eqref{eq:mse}, we have 
\begin{align}
	e_{k, \mathbf{f}_{k, \star}}&=P\mathbf{f}_{k,\star}^{\textsf{H}}\mathbf{F}_{\mathrm{RF}}^{\textsf{H}}\left( \sum\nolimits_{j=1}^K{\mathbf{h}_j\mathbf{h}_{j}^{\textsf{H}}}+\sigma ^2\mathbf{I}_M \right) \mathbf{F}_{\mathrm{RF}}^{}\mathbf{f}_{k,\star}^{}\notag \\
	&-2P\Re \left\{ \mathbf{f}_{k,\star}^{\textsf{H}}\mathbf{F}_{\mathrm{RF}}^{\textsf{H}}\mathbf{h}_k \right\} +P .
\end{align}
Therefore, the sub-problem with respect to $\mathbf{F}_{\rm RF}$ can be written in the following form:
\begin{problem}\label{pd:uplink_sum_rate_2} 
	\begin{align}
		\underset{\mathbf{F}_{\rm RF}}{\min}\quad & \sum\nolimits_{k=1}^K { \omega _k e_{k, \mathbf{f}_{k, \star}} }  \quad \text{s.t.}\quad \eqref{1tst:3}. \label{obj_func:mse_Wrf}
	\end{align}
\end{problem}
Furthermore, defining $\mathbf{R}\triangleq \sum\nolimits_{j=1}^K{\mathbf{h}_j\mathbf{h}_{j}^{\textsf{H}}}+(\sigma ^2/P)\mathbf{I}_M$, $\mathbf{C}\triangleq P\sum\nolimits_{k=1}^K{\omega_k\mathbf{f}_{k,\star}^{}\mathbf{f}_{k,\star}^{\textsf{H}}}$, and $\mathbf{B}\triangleq 2P\sum\nolimits_{k=1}^K{\omega_k\mathbf{h}_k\mathbf{f}_{k,\star}^{\textsf{H}}}$, the MSE expression can be re-written in a more compact form:
\begin{align}
	&\sum\nolimits_{k=1}^K{\omega_k e_{k,\mathbf{f}_{k,\star}}}=P\sum\nolimits_{k=1}^K{\omega_k\mathrm{tr}\left\{ \mathbf{F}_{\mathrm{RF}}^{\textsf{H}}\mathbf{RF}_{\mathrm{RF}}^{}\mathbf{w}_{k,\star}^{}\mathbf{w}_{k,\star}^{\textsf{H}} \right\}}\notag\\
	&\quad-2P\sum\nolimits_{k=1}^K{\Re \left\{ \mathrm{tr}\left\{ \mathbf{F}_{\mathrm{RF}}^{\textsf{H}}\mathbf{h}_k\mathbf{f}_{k,\star}^{\textsf{H}} \right\} \right\}} + KP \notag \\
	&\quad=P\mathrm{tr}\left\{ \mathbf{F}_{\mathrm{RF}}^{\textsf{H}}\mathbf{RF}_{\mathrm{RF}}^{}\sum\nolimits_{k=1}^K{\mathbf{f}_{k,\star}^{}\mathbf{f}_{k,\star}^{\textsf{H}}} \right\} \notag \\
	&\quad-2P\Re \left\{ \mathrm{tr}\left\{ \mathbf{F}_{\mathrm{RF}}^{\textsf{H}}\sum\nolimits_{k=1}^K{\mathbf{h}_k\mathbf{f}_{k,\star}^{\textsf{H}}} \right\} \right\} + KP\notag \\
	&\quad=\mathrm{tr}\left\{ \mathbf{F}_{\mathrm{RF}}^{\textsf{H}}\mathbf{RF}_{\mathrm{RF}}^{}\mathbf{C} \right\} -\Re \left\{ \mathrm{tr}\left\{ \mathbf{F}_{\mathrm{RF}}^{\textsf{H}}\mathbf{B} \right\} \right\} +KP. \label{eq:sum_error}
\end{align}
Therefore, to minimize the above equation under the unit-modulus constraint, we need to align the phase terms in $\mathbf{F}_{\rm RF}$ with $\mathbf{R}$, $\mathbf{B}$, and $\mathbf{C}$.
Hence, we adopt an element-wise optimization strategy for the PS phases. 
Specifically, we update the entries of $\mathbf{F}_{\rm RF}$ sequentially, i.e., we optimize one entry at a time while keeping the remaining entries fixed.

To facilitate the narrative, we define the entry-selection matrix $\mathbf{E}_{m,n}$ by setting the $(m,n)$-th entry to one and all other entries to zeros. 
Therefore, to update the $(m,n)$-th entry of $\mathbf{F}_{\rm RF}$, the analog beamforming matrix can be written in the following form
\begin{align}
    \mathbf{F}_{\rm RF} = \breve{\mathbf{F}}_{m,n} + w \mathbf{E}_{m,n}, \label{eq:decompose}
\end{align}
where $\breve{\mathbf{F}}_{m,n}$ is obtained by setting the $(m,n)$-th entry of $\mathbf{F}_{\rm RF}$ as zero while keeping the rest entries unchanged, and $w$ stands for the optimization variable $[\mathbf{F}_{\rm RF}]_{m,n}$.
The PC structure requires a lower number of PSs compared to the FC structure.
Therefore, the element-wise phase calibration in \eqref{pd:uplink_sum_rate_3} needs to be applied only to the nonzero parameters, thereby making RF beamforming less complex as a result of the sparse structure.
In addition, the sparse PC structure enables a more stable phase optimization than the FC structure, since each phase is coupled with a smaller set of other phases.
Plugging \eqref{eq:decompose} into the MSE expression, the MSE expression can be rewritten as an element-wise form, which is given by
\begin{align}
    &\sum\nolimits_{k=1}^K{\omega_k e_{k,\mathbf{f}_{k,\star}}} = |w|^2\mathrm{tr}\left\{ \mathbf{E}_{m,n}^{\textsf{H}}\mathbf{RE}_{m,n}\mathbf{C} \right\} \notag \\
    &+\Re \left\{ \mathrm{tr}\left\{ w^*\left( 2\mathbf{E}_{m,n}^{\textsf{H}}\mathbf{R}\breve{\mathbf{F}}_{m,n}\mathbf{C}-\mathbf{E}_{m,n}^{\textsf{H}}\mathbf{B} \right) \right\} \right\} +c_{1, m,n}, \notag
\end{align}
where the constant term is defined as $c_{1, m,n}\triangleq \mathrm{tr}\{ \breve{\mathbf{F}}_{m,n}^{\textsf{H}}\mathbf{R}\breve{\mathbf{F}}_{m,n}\mathbf{C} \} -\Re \{ \mathrm{tr}\{ \breve{\mathbf{F}}_{m,n}^{\textsf{H}}\mathbf{B} \} \} +KP$.
Further, it is convenient to introduce the following definitions:
\begin{align}
    c_{2,m,n}&\triangleq \mathrm{tr}\left\{ \mathbf{E}_{m,n}^{\textsf{H}}\mathbf{RE}_{m,n}\mathbf{C} \right\}. 
    \\
    c_{3,m,n}&\triangleq \mathrm{tr}\left\{\mathbf{E}_{m,n}^{\textsf{H}}\mathbf{B}\right\}-2\mathrm{tr}\left\{\mathbf{E}_{m,n}^{\textsf{H}}\mathbf{R}\breve{\mathbf{F}}_{m,n}\mathbf{C}\right\}.
\end{align}
Based on these definitions, the optimization problem w.r.t. the $(m,n)$-th entry of $\mathbf{F}_{\rm RF}$ can be formulated as follows:
\begin{problem}\label{pd:uplink_sum_rate_3} 
	\begin{align}
		\underset{w}{\min}\quad & c_{2,m,n}|w|^2-\Re \left\{ \mathrm{tr}\left\{ w^*c_{3,m,n} \right\} \right\} +c_{1,m,n} \label{obj_func:mse_Wrf_element} \\
		\text{s.t.}\quad & |w|=1. 
	\end{align}
\end{problem}
Therefore, the optimal solution to problem \eqref{pd:uplink_sum_rate_3} can be obtained as follows:
\begin{align}
	w_{i,j,\star} = \mathrm{e}^{\mathrm{j}\angle c_{3, m,n}}. \label{eq:update_PSs}
\end{align}
Applying \eqref{eq:update_PSs} to all entries in $\mathbf{F}_{\rm RF}$, the optimized analog beamforming matrix denoted by $\mathbf{F}_{\rm RF, \star}$ can be obtained as $[\mathbf{F}_{\rm RF, \star}]_{m,n}=w_{i,j,\star}$.
The pinching beamforming is performed again using the Gauss-Seidel approach in \textbf{Algorithm \ref{alg:element_wise}}.
In addition to the above steps, the auxiliary variable set $\Omega$ is updated according to \eqref{eq:auxiliary_update}.
Thus, the overall BCD framework follows \textbf{Algorithm~\ref{alg:overall}}, with the only difference being that the Riemannian gradient descent block is replaced by the proposed element-wise phase optimization.
The computational complexity can be derived as $\mathcal{O}(I_{\rm iter, 1} (M^3 + N +  I_{\rm iter 3}M + K)$, where $I_{\rm iter 3}$ is attributed to the element-wise phase optimization.

\section{Performance Analysis of SWAN-based Hybrid Beamforming Framework} \label{sect:preformance_analysis}
To characterize how the achievable throughput scales with the number of segments, we consider two scenarios: 1) A single RF chain with multiple PSs, which is a special case of the FC structure; and 2) multiple RF chains with a single PA, which is a special case of the PC structure.
To make the analysis tractable, we consider the single-user case.

\subsection{Fully-Connected Limit: Single RF Chain with Multiple PSs} \label{sect:sub-performance_analysis_case1}
In this case, the optimization variable is a complex-valued scalar, i.e., $|w|^2=1$, and an analog beamforming vector, i.e., $\mathbf{w}_{\rm RF} \in \mathbb{C}^{M \times 1}$. 
Hence, the optimization problem can be simplified into the following form:
\begin{problem}\label{pd:performance_analysis} 
	\begin{align}
		\underset{\mathbf{w}_{\rm RF}, \mathbf{x}}{\max} 
		\quad & \log _2\left( 1+P\left| \mathbf{w}_{\mathrm{RF}}^{\textsf{H}}\mathbf{h}\left( \mathbf{x} \right) \right|^2/(M\sigma ^2) \right)  \\
		\text{s.t.}\quad 
		& \left| [\mathbf{w}_{\mathrm{RF}}]_{i} \right|=1,\ \forall i,\quad  \mathbf{x} \in \mathcal{F}, 
	\end{align}
\end{problem}
where we deliberately dropped the user index for simplicity. 
Therefore, the optimal solution for $\mathbf{w}_{\rm RF}$ under the unit modulus constraint can be derived as $\mathbf{w}_{\rm RF, \star} = \mathrm{e}^{\mathrm{j}\angle \mathbf{h}(\mathbf{x})}$.
Therefore, plugging the expression of $\mathbf{w}_{\rm RF, \star}$ into the objective function, the original optimization problem can be rewritten as $\mathbf{x}_{\star}=\arg\max_{\mathbf{x}\in \mathcal{F}}
(\sum\nolimits_{m=1}^{M} |[\mathbf h(\mathbf x)]_m |)^2$, which is an optimization problem w.r.t. the PA positions $\mathbf{x}$.
Moreover, the objective function is a monotonically increasing function w.r.t. the channel gain.
Based on the above and the channel vector expression in \eqref{eq:overall_channel}, the channel gain is given by
\begin{align}
    \left(\sum\nolimits_{m=1}^{M} \big|[\mathbf h(\mathbf x)]_m\big|\right)^2 \overset{(a)}{\simeq} \eta \left| \sum\nolimits_{m=1}^M{1/r_m} \right|^2, \label{eq:channel_power}
\end{align}
where $(a)$ is achieved by omitting the in-waveguide losses in each segment.
It is noted that, as proved in \cite{ouyang2025uplink}, the in-waveguide loss is negligible for a moderate number of segments.
Specifically, a channel-gain loss of less than $10~\%$ can be guaranteed for $M\ge 9$ over a total waveguide length of $100$~meters.
Hence, the assumption of negligible in-waveguide loss is mild in our experimental setup.
Thus, the optimal PA positions are obtained by placing the PAs as close to the user as possible to minimize the path loss in each link.
In particular, the first step is to select the segment indexed by $m_\star$ that contains the user's $x$-coordinate and place the PA on this segment at the user's $x$-coordinate, i.e., ${x}_{m_\star} = r_x$.
Subsequently, subject to the inter-spacing constraints between PAs and the coordinate range of each segment, the next PA position on the right-hand side, specifically the $(m_\star+1)$-th segment, is determined by
\begin{align}
    {x}_{m_\star+1} = \max \{ {x}_{m_\star} + \Delta_{\min}, x_{m_\star+1}^{\rm FD}\}.
\end{align}
By applying this method sequentially to the rest of the PAs, the optimal PA position vector denoted by $\mathbf{x}_{\star}$ is obtained.
Consequently, the resulting maximum channel gain is given as follows:
\begin{align}
    &\left| \mathbf{h}(\mathbf{x}_{\star}) \right|^2= \eta \left| \sum\nolimits_{m=1}^M{\left( 1/\sqrt{\left( {x}_{m^{\star}}-r_x \right) ^2+r_{y}^{2}+H^2} \right)} \right|^2.
\end{align}
Assuming that the target is located at the center of the service area, the maximum SNR can be obtained as follows:
\begin{align}
    \gamma _{\max}^{(1)}=\frac{P\eta}{M\sigma ^2}\left( \frac{1}{\Delta _{yz}}+\sum\nolimits_{\tilde{m}=1}^{\frac{M-1}{2}}{\frac{2}{\sqrt{L^2\left( \tilde{m}-0.5 \right) ^2+\Delta _{yz}^{2}}}} \right) ^2,
\end{align}
where $\Delta _{yz}\triangleq (r_{y}^{2}+H^2)^{1/2}$.
According to the Euler-Maclaurin formula and $(M-1)L \simeq ML = D_x$ \cite{ouyang2025uplink}, for large $M$, the above equation can be approximated by the following expression:
\begin{align}
    \gamma_{\max}^{(1)} \simeq \frac{P\eta}{M\sigma^2} \left( \frac{1}{\Delta _{yz}}+\frac{2}{L}\sinh ^{-1}\left( \frac{LM}{2\Delta _{yz}^{}} \right) \right) ^2. \label{eq:g_max_case_1}
\end{align}
Hence, the rate is given by
\begin{align}
    R_{\max}^{(1)} = \log_2 (1 + \gamma_{\max}^{(1)}). \label{eq:rate_case_1}
\end{align}
\begin{remark} \label{remark_1}
(Rate Scaling Law w.r.t. Number of Segments $M$)
\emph{Based on \eqref{eq:g_max_case_1} and \eqref{eq:rate_case_1}, the rate is a non-monotonic function w.r.t. the number of segments $M$ for a given segment length $L$. 
More specifically, the rate first increases w.r.t. $M$ until $M_\star = 2\Delta_{yz}^2/L$; after this point, the rate drops as $M$ increases further.
The reason is twofold: 1) As the number of segments increases, the aggregated noise power, i.e., $M\sigma^2$, increases accordingly, which reduces the SNR; and 2) For a fixed total length, additional segments are progressively farther from the user, yielding diminishing contributions to the beam gain enhancement.
Consequently, the increased noise power offsets the beamforming gain, resulting in a reduced achievable rate.
    For sufficiently large $M$, we have $R_{\max}^{(1)} = \mathcal{O}((\log M)^2/M)$ and $R_{\max}^{(1)}\to 0$ as $M\to\infty$.}
\end{remark}

\subsection{Partially-Connected Limit: Multiple RF Chains with Single PS}
In this part, we consider the single-user scenario where the number of RF chains equals the number of segments, and each chain is connected to a feed point via a PS.
In this case, since the phase shifts and amplitude tuning can be freely adjusted by the digital beamforming vector $\|\mathbf{w}\|=1$, we can discard the optimization of the analog beamforming vector $\mathbf{w}_{\rm RF}$ by regarding it as an inherent part of the channel vector.
Without loss of generality, we set the phases of the PSs to zero.
Consequently, the optimization problem is formulated as follows: 
\begin{problem}
	\begin{align}
		  \underset{\mathbf{w},\mathbf{x}}{\max}\quad &\log _2(1+|\mathbf{w}^{\textsf{H}}\mathbf{h}\left( \mathbf{x} \right) |^2/(\sigma ^2\|\mathbf{w}\|^2))\\
		\text{s.t.}\quad 
		& \mathbf{x} \in \mathcal{F}.
	\end{align}
\end{problem}
For a fixed PA position, the optimal beamforming vector follows the maximum-ratio combining (MRC) rule, i.e., $\mathbf{w}_{\star}=\frac{\mathbf{h}(\mathbf{x})}{\parallel \mathbf{h}(\mathbf{x})\parallel}$, whose optimality directly follows from the Cauchy-Schwarz inequality.
Consequently, this problem can be equivalently written as $\mathbf{x}_{\star}=\max_{\mathbf{x}\in \mathcal{F}} \left\| \mathbf{h}\left( \mathbf{x} \right) \right\| ^2$, where the channel gain can be expressed as follows:
\begin{align}
    \left\| \mathbf{h}(\mathbf{x}) \right\| ^2\overset{(a)}{\simeq}\eta \sum\nolimits_{m=1}^M{1/r_{m}^{2}}.
\end{align}
Here, we again omit the in-waveguide loss.
As in the former scenario discussed in Section \ref{sect:sub-performance_analysis_case1}, the optimal PA positions, i.e., $\mathbf{x}_{\star}$, are obtained by placing the PAs as close to the target as possible, since the path loss is minimized.
Therefore, the maximized channel gain is given by 
\begin{align}
    \gamma _{\max}^{(2)}=\frac{P\eta}{\sigma ^2}\left( \frac{1}{\Delta _{yz}^{2}}+\sum\nolimits_{\tilde{m}=1}^{\frac{M-1}{2}}{\frac{2}{L^2\left( \tilde{m}-0.5 \right) ^2+\Delta _{yz}^{2}}} \right). \label{eq:snr_case_2}
\end{align}
For a sufficiently large odd $M$, this equation can be approximated by 
\begin{align}
    \gamma _{\max}^{(2)}\simeq\frac{P\eta}{\sigma ^2}\left( \frac{1}{\Delta _{yz}^{2}}+\frac{2}{L\Delta _{yz}^{}}\tan ^{-1}\left( \frac{\left( M-1 \right) L}{2\Delta _{yz}^{}} \right) \right). \label{eq:snr_case_2}
\end{align}
Correspondingly, the rate is given by 
\begin{align}
     R_{\max}^{(2)} = \log_2 (1 + \gamma_{\max}^{(2)}). \label{eq:rate_case_2}
\end{align}
\begin{remark}
(Rate Scaling Law w.r.t. Number of Segments $M$)
\emph{Based on \eqref{eq:snr_case_2} and \eqref{eq:rate_case_2}, the rate is monotonically increasing with $M$ but bounded for fixed segment length $L$.
The rate increases with increasing $M$ and converges to $\gamma _{\max}^{(2, M\rightarrow \infty)}=\frac{P\eta}{\sigma ^2}\left( \frac{1}{\Delta _{yz}^{2}}+\frac{\pi}{L\Delta _{yz}^{}} \right)$.
Compared with the former case in Remark \ref{remark_1}, where the achievable rate decreases with $M$, the bounded rate-scaling law in this case stems from the fact that MRC beamforming aggregates noise noncoherently, such that the effective noise power does not scale with the number of segments.
Consequently, as the additional segments become increasingly distant from the user's position, the incremental rate gain from these segments diminishes due to greater path loss, causing the rate to converge to a fixed value.
For sufficiently large $M$, we have $R_{\max}^{(2,M\rightarrow \infty )}=\lim_{M\rightarrow \infty} R_{\max}^{(2)}=\log _2\left( \frac{P\eta}{\sigma ^2}\left( \frac{1}{\Delta _{yz}^{2}}+\frac{\pi}{L\Delta _{yz}^{}} \right) \right)$.
In other words, the rate is bounded, i.e., $R_{\max}^{(2,M\rightarrow \infty )}\propto \mathcal{O} \left( 1 \right)$.}
\end{remark}

\section{Numerical Results} \label{sect:results}
The parameters in Table \ref{tab:sim_params} are used throughout our simulations unless otherwise specified.
\begin{table}[t]
\centering
\caption{Simulation Parameters.}
\label{tab:sim_params}
\footnotesize
\renewcommand{\arraystretch}{1.05}
\setlength{\tabcolsep}{6pt}
\begin{tabular}{l c}
\hline
\textbf{Parameter (Symbol)} & \textbf{Value} \\
\hline
Carrier frequency ($f$) & $28~\mathrm{GHz}$ \\
Effective refractive index ($n_{\rm eff}$) & $1.4$ \\
In-waveguide loss ($\kappa$) & $0.08~\mathrm{dB/m}$ \\
Transmit power ($P$) & $10~\mathrm{dBm}$ \\
Noise power ($\sigma^2$) & $-80~\mathrm{dBm}$ \\
Service area size ($D_x \times D_y$) & $80~\mathrm{m}\times 20~\mathrm{m}$ \\
Receiver height ($H$) & $3~\mathrm{m}$ \\
Number of users ($K$) & $4$ (uniformly distributed) \\
Number of RF chains (tri-hybrid) ($N_{\rm RF}$) & $25$ \\
Number of segments (tri-hybrid) ($M$) & $50$ \\
Number of pinching antennas ($N_{\rm PA}$) & $50$ \\
Gauss--Seidel search resolution & $0.01~\mathrm{m}$ \\
Minimum PA inter-spacing ($\Delta_{\min}$) & $\lambda_{\rm c}/2$ \\
BCD stopping criterion ($\epsilon$) & fractional decrease $<10^{-8}$ \\
\hline
\end{tabular}
\vspace{-1em}
\end{table}
Furthermore, the WMMSE-based method utilizes the beamformers obtained with the ZF-based algorithm for initialization.
The numerical results presented in this section are based on 1000 Monte Carlo trials.
The following benchmarks are considered to validate the effectiveness of the proposed methods:
\begin{itemize}
    \item \textbf{mMIMO with an FC structure and WMMSE (mMIMO, FC, WMMSE)}:
    This benchmark uses a conventional mMIMO receiver to serve multiple users. 
    The receiver is equipped with $M=50$ antennas and $N_{\rm RF}=25$ RF chains, and the array is located at the center of the service area. The optimization procedure for this baseline follows the WMMSE-based method in Section~\ref{sect:mmse_fully}. Unlike the proposed tri-hybrid design, this benchmark removes the pinching beamforming step, thereby isolating and highlighting the benefits of tri-hybrid beamforming.
    
    \item \textbf{Conventional PASS without IAR (Conv. PASS w/o IAR)}:
    This benchmark uses a conventional PASS receiver to serve multiple users. Unlike the SWAN-based architecture, PASS employs a single long waveguide spanning the service area along the $x$-direction. The number of pinching antennas is set to $M=50$, while the number of RF chains is limited to $N_{\rm RF}=1$ due to the constraints of conventional PASS. 
    More importantly, we focus on an ideal scenario in which the IAR effect is not taken into account.
\end{itemize}
Alongside the baseline schemes, ``SWAN FC, WMMSE” and ``SWAN PC, WMMSE” refer to the proposed SWAN-based tri-hybrid WMMSE beamforming designs for the FC and PC structures, respectively. ``SWAN FC, ZF” stands for the proposed SWAN-based tri-hybrid ZF beamforming design under the FC architecture.

\subsection{Convergence Behavior and Impact of Transmit Power}
\begin{figure}[t!]
    \centering
    \includegraphics[width=0.8\linewidth]{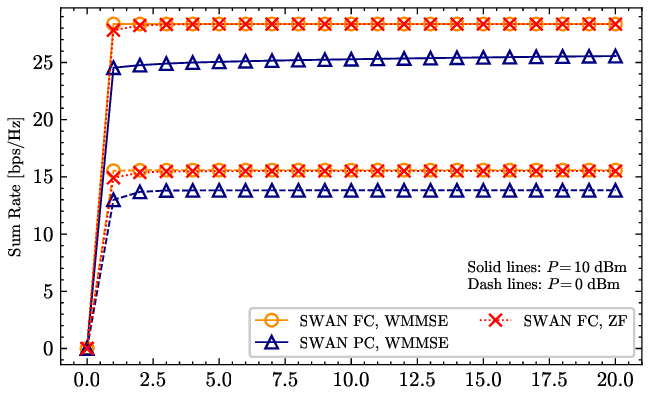}
    \caption{Illustration of the convergence behavior at $P=0~\mathrm{dBm}$ and $P=10~\mathrm{dBm}$.}
    \label{fig:convergence_behavior}
\end{figure}
Fig. \ref{fig:convergence_behavior} shows the convergence behavior of the proposed algorithms for different transmit powers.
It is observed that the proposed algorithms converge quickly within 10 iterations, which indicates the effectiveness of the proposed methods.
Moreover, due to the DoF loss incurred by the PC structure, there is a performance gap between the ``SWAN PC, WMMSE" and ``SWAN FC, WMMSE" results.
In the considered uplink hybrid beamforming architecture, the analog combiner restricts the effective DoFs available to the digital combiner. 
As such, the additional optimization gain offered by WMMSE over ZF becomes marginal, yielding nearly identical performance.

\begin{figure}[t!]
    \centering
    \includegraphics[width=0.8\linewidth]{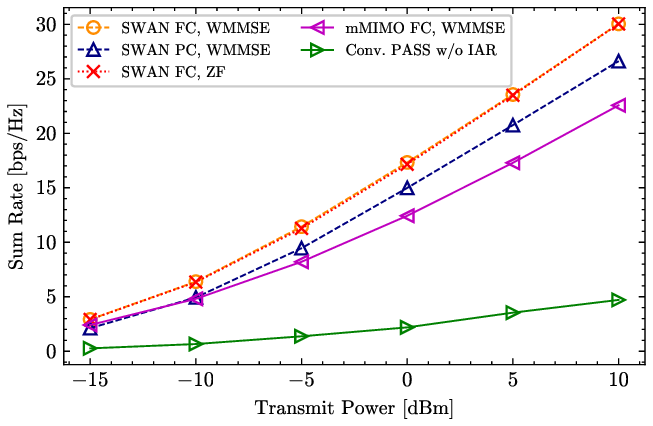}
    \caption{Illustration of sum rate versus transmit power.\vspace{-1em}}
    \label{fig:sum_rate_vs_transmit_power}
\end{figure}
In Fig. \ref{fig:sum_rate_vs_transmit_power}, we investigate the sum rate as a function of transmit power.
As the transmit power increases, the sum rate increases correspondingly.
Due to the lack of pinching beamforming, i.e., the third layer of the tri-hybrid beamforming architecture, the ``mMIMO FC, WMMSE" baseline yields a worse performance than the proposed methods.
Furthermore, the multi-user interference (MUI) cannot be effectively suppressed in the ``Conv. PASS" baseline, due to the limited number of RF chains.
Consequently, it performs poorly in the considered multi-user scenario.

\subsection{Impact of Number of RF Chains}

\begin{figure}[t!]
    \centering
    \includegraphics[width=0.8\linewidth]{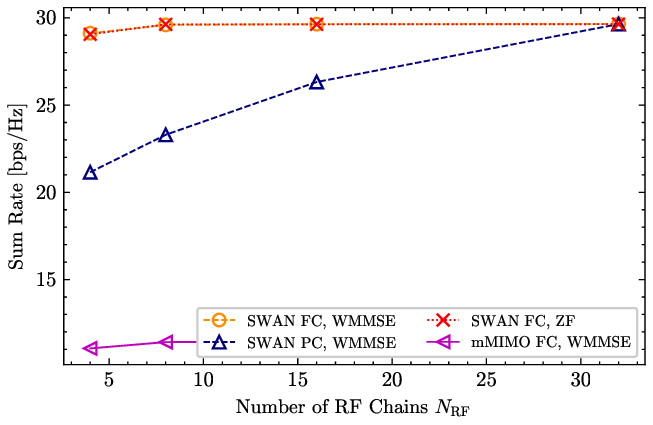}
    \caption{Illustration of sum rate versus the number of RF chains.\vspace{-1em}}
    \label{fig:sum_rate_vs_num_of_rf_chain}
\end{figure}

\begin{figure}[t!]
    \centering
    \includegraphics[width=0.8\linewidth]{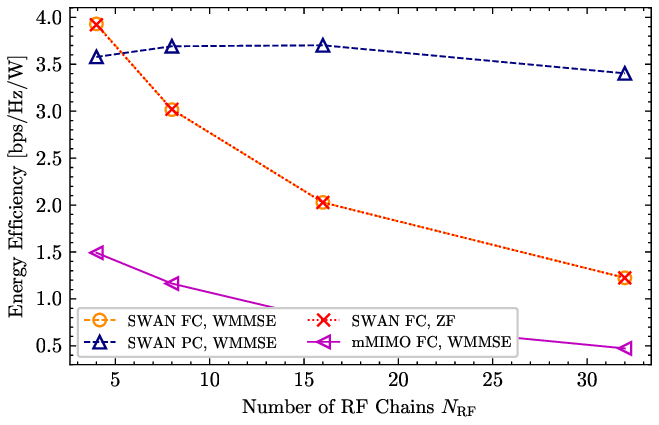}
    \caption{Illustration of energy efficiency versus the number of RF chains.\vspace{-1em}}
    \label{fig:ee_vs_transmit_power}
\end{figure}
Figs. \ref{fig:sum_rate_vs_num_of_rf_chain} and \ref{fig:ee_vs_transmit_power} illustrate how the sum rate and energy efficiency (EE) vary with an increasing number of RF chains.
In this subsection, we set the number of segments to $M=32$ to guarantee that the number of segments in one sub-array in the PC structure, i.e., $N_{\rm par}$, is an integer.
Moreover, to evaluate EE, we adopt the energy consumption model in \cite{yu2016alternating, jiang2025beam}.
In particular, let the energy consumption of one power amplifier, one phase shifter, and one RF chain be $P_{\rm PA}=100~\mathrm{mW}$, $P_{\rm PS}=10~\mathrm{mW}$, and $P_{\rm RF}=100~\mathrm{mW}$, respectively.
The energy efficiency is defined as $\mathrm{EE}=\frac{R}{P+N_{\mathrm{RF}}P_{\mathrm{RF}}+MP_{\mathrm{PA}}+N_{\mathrm{PS}}P_{\mathrm{PS}}}$, where $N_{\mathrm{PS}}$ denotes the total number of PSs, which is given by $N_{\mathrm{PS}}=M \times N_{\rm RF}$ for the FC structure and $N_{\mathrm{PS}}=M$ for the PC structure.

As shown in Fig. \ref{fig:sum_rate_vs_num_of_rf_chain}, the sum rate increases with an increasing number of RF chains, as a result of the enhancement of the interference suppression.
In addition, due to partial connectivity, the proposed ``SWAN PC, WMMSE" has a performance gap compared to the other approaches.
Furthermore, the conventional mMIMO baseline performs worst, underscoring the benefits of pinching beamforming.
In terms of EE, as shown in Fig. \ref{fig:ee_vs_transmit_power}, due to the large number of segments and the small transmit power, the EE of the FC structure decreases as $N_{\rm RF}$ increases.
In contrast, the partial connectivity can reduce the hardware energy consumption by enforcing a sparse analog beamforming matrix, yielding a performance gain compared to the FC structure as $N_{\rm RF}$ increases.
\vspace{-1em}
\subsection{Impact of Number of Users}
\begin{figure}[t!]
    \centering
    \includegraphics[width=0.8\linewidth]{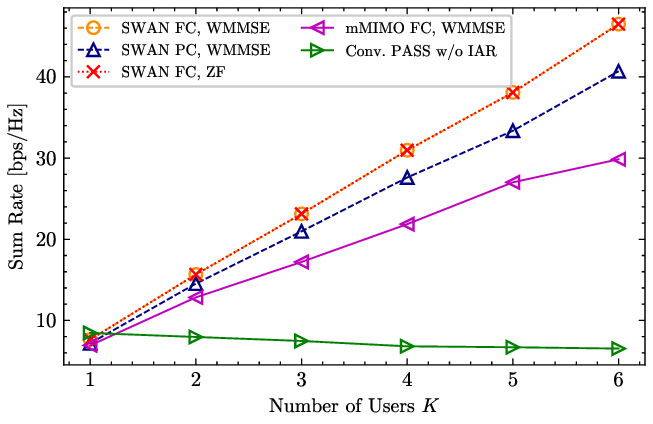}
    \caption{Illustration of sum rate versus user numbers.\vspace{-1em}}
    \label{fig:sum_rate_vs_user_num}
\end{figure}
Fig. \ref{fig:sum_rate_vs_user_num} shows the sum rate versus an increasing number of users.
It can be observed that, as the number of users increases, the achievable sum rate increases simultaneously.
However, as the number of users increases, the gap between the PC and FC structures widens.
This is because, as the number of users increases, the reduced DoFs of the sparse analog beamforming matrix limit the system's overall performance.
As for the baselines, the mMIMO framework shows a performance gap relative to the proposed methods due to its lack of channel reconfigurability.
Conventional PASS has limited performance due to its single-RF-chain structure.

\subsection{Impact of Side Lengths}
\begin{figure}[t!]
    \centering
    \includegraphics[width=0.8\linewidth]{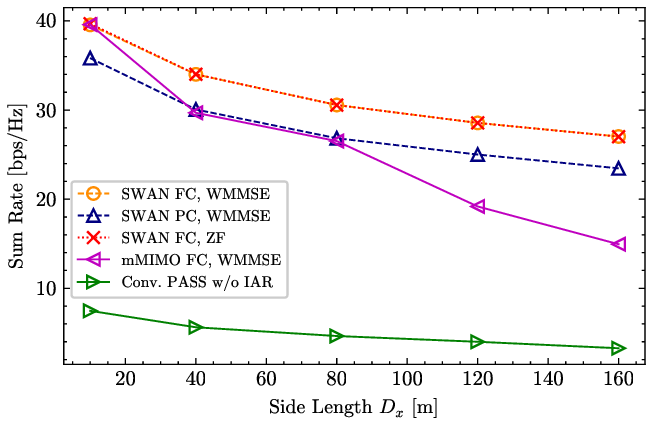}
    \caption{Illustration of sum rate versus side length.\vspace{-1em}}
    \label{fig:sum_rate_vs_side_length}
\end{figure}
Fig. \ref{fig:sum_rate_vs_side_length} shows the achievable sum rate as a function of the side length of the service area.
In particular, we enlarge the side length in the $x$-direction while keeping the side length in the $y$-direction unchanged.
In addition, corresponding to an increasing side length, the total length of the SWAN receiver increases for a fixed number of segments $M$.
As the side length increases, the achievable sum rate decreases.
Since the number of segments remains unchanged, the number of RF chains per unit length decreases, indicating a reduction in signal processing capability.
However, compared with the conventional mMIMO baseline, the reduction ratio in the achievable sum rate is more robust for the SWAN receiver.

\subsection{Rate Scaling Law for The Number of Segments}
\begin{figure}[t!]
    \centering
    \includegraphics[width=0.83\linewidth]{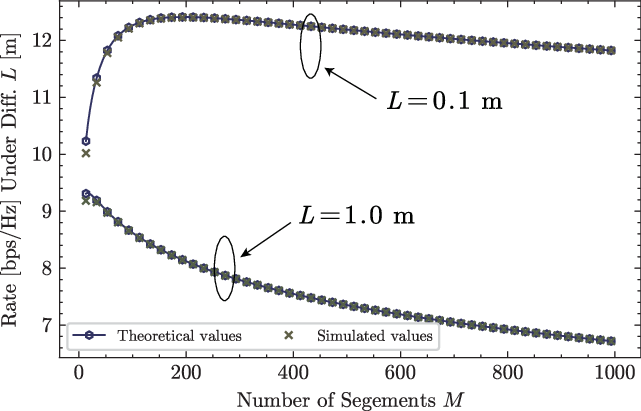}
    \caption{Illustration of FC limit for a single RF chain and multiple PSs.\vspace{-1em}}
    \label{fig:performance_fully_connected}
\end{figure}

In this section, we verify the derived rate-scaling laws for both the FC and PC structures.
To this end, as mentioned in Section \ref{sect:preformance_analysis}, the single-user case is considered. 
In particular, Fig. \ref{fig:performance_fully_connected} shows the rate-scaling laws for the FC structure with a single RF chain and multiple PAs.
As shown in this figure, the theoretical derivations match the simulation results with high accuracy.
As the number of segments increases, the achievable rate exhibits an increasing-then-decreasing behavior, which is aligned well with our analysis in Section \ref{sect:preformance_analysis}.
For a given number of segments, the achievable rate decreases with increasing segment length, because more segments at both ends of the SWAN are farther away from the user's location.
In addition, the optimal number of segments decreases as the length of each segment increases, indicating that the contribution of segments distant from the user diminishes.

\begin{figure}[t!]
    \centering
    \includegraphics[width=0.83\linewidth]{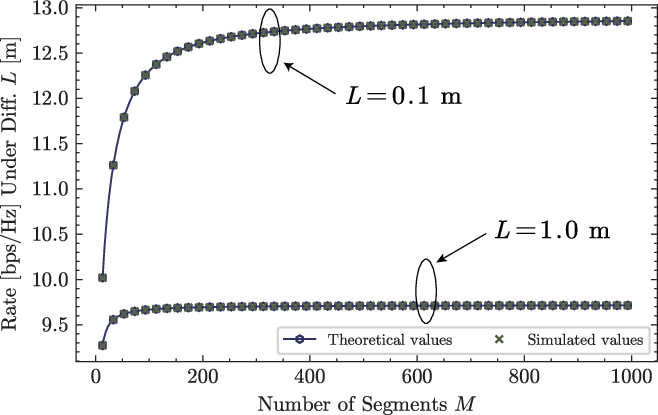}
    \caption{Illustration of PC limit for multiple RF chains and a single PS.\vspace{-1em}}
    \label{fig:performance_partially_connected}
\end{figure}
Fig. \ref{fig:performance_partially_connected} shows the rate-scaling law for the PC structure with multiple RF chains and a single PS.
As the number of segments increases, the rate increases but remains bounded by the respective asymptotic value.
Moreover, as the segment length increases, the rate drops because more distant segments contribute less to the overall rate maximization.
Finally, the agreement between theoretical and simulation results confirms the correctness of our derivations.

\section{Conclusions} \label{sect:conclusion}
This paper investigated SWAN-based tri-hybrid beamforming for uplink multi-user MIMO sum-rate maximization under both the FC and PC topologies.
For the FC structure, WMMSE- and ZF-based tri-hybrid beamforming schemes were developed for uplink sum-rate maximization, where a BCD framework was employed for optimization of digital beamforming, Riemannian-manifold optimization of analog beamforming, and Gauss-Seidel-based PA repositioning.
For the PC structure, an interleaved topology was proposed for SWAN, and a WMMSE-based tri-hybrid design was devised.
To leverage the sparse analog architecture of the PC structure, an element-wise phase calibration was adopted to reduce complexity within the BCD framework.
We further derived the rate-scaling law as a function of the number of segments, assuming simplified scenarios for both the FC and PC structures.
The analytical results showed that increasing the number of segments does not necessarily improve the achievable rate.
Extensive numerical results verified the analysis and demonstrated that the proposed SWAN-based tri-hybrid beamforming consistently outperforms conventional mMIMO with hybrid beamforming and conventional PASS with pinching-only beamforming for both the FC and PC structures.

\bibliographystyle{IEEEtran}
\bibliography{refs}

@ARTICLE{shafi20175g,
  author={Shafi, Mansoor and others},
  journal={IEEE J. Sel. Areas Commun.}, 
  title={{5G}: A Tutorial Overview of Standards, Trials, Challenges, Deployment, and Practice}, 
  year={Jun. 2017},
  volume={35},
  number={6},
  pages={1201-1221},
  doi={10.1109/JSAC.2017.2692307}}

@ARTICLE{ning2026precoding,
  author={Ning, Boyu and others},
  journal={IEEE Commun. Surveys Tuts.}, 
  title={Precoding Matrix Indicator in the {5G NR} Protocol: A Tutorial on {3GPP} Beamforming Codebooks}, 
  year={2026, early access, doi:10.1109/COMST.2026.3653568},
  volume={},
  number={},
  pages={1-1},
  doi={10.1109/COMST.2026.3653568}}

@ARTICLE{liu2025near,
  author={Liu, Yuanwei and others},
  journal={IEEE Commun. Surveys Tuts.}, 
  title={Near-Field Communications: A Comprehensive Survey}, 
  year={Jun. 2025},
  volume={27},
  number={3},
  pages={1687-1728},
  doi={10.1109/COMST.2024.3475884}}

@article{björnson2026antenna,
      title={From Antenna Abundance to Antenna Intelligence in {6G} Gigantic {MIMO} Systems}, 
      author={Emil Björnson and others},
      year={2026},
      journal={arXiv preprint arXiv:2601.08326}
}

@ARTICLE{gong2024holographic,
  author={Gong, Tierui and others},
  journal={IEEE Commun. Surveys Tuts.}, 
  title={Holographic {MIMO} Communications: Theoretical Foundations, Enabling Technologies, and Future Directions}, 
  year={1st Quart. 2024},
  volume={26},
  number={1},
  pages={196-257},
  doi={10.1109/COMST.2023.3309529}}

@ARTICLE{lu2014overviewmimo,
  author={Lu, Lu and Li, Geoffrey Ye and Swindlehurst, A. Lee and Ashikhmin, Alexei and Zhang, Rui},
  journal={IEEE J. Sel. Topics Signal Process.}, 
  title={An Overview of Massive {MIMO}: Benefits and Challenges}, 
  year={Oct. 2014},
  volume={8},
  number={5},
  pages={742-758},
  doi={10.1109/JSTSP.2014.2317671}}

@ARTICLE{liu2025pinchingantenna,
  author={Liu, Yuanwei and Jiang, Hao and Xu, Xiaoxia and others},
  journal={IEEE Trans. Commun.}, 
  title={Pinching-Antenna Systems ({PASS}): A Tutorial}, 
  year={Jan. 2026},
  volume={74},
  number={},
  pages={4881-4918},
  doi={10.1109/TCOMM.2026.3658289}}

@ARTICLE{new2025fulid_tutorial,
  author={New, Wee Kiat and Wong, Kai-Kit and Xu, Hao and others},
  journal={IEEE Commun. Surveys Tuts.}, 
  title={A Tutorial on Fluid Antenna System for {6G} Networks: Encompassing Communication Theory, Optimization Methods and Hardware Designs}, 
  year={Aug. 2025},
  volume={27},
  number={4},
  pages={2325-2377},
  doi={10.1109/COMST.2024.3498855}}

@ARTICLE{zhu2024movable,
  author={Zhu, Lipeng and Ma, Wenyan and Zhang, Rui},
  journal={IEEE Commun. Mag.}, 
  title={Movable Antennas for Wireless Communication: Opportunities and Challenges}, 
  year={Jun. 2024},
  volume={62},
  number={6},
  pages={114-120},
  doi={10.1109/MCOM.001.2300212}}

@ARTICLE{alkhateeb2014mimo,
  author={Alkhateeb, Ahmed and others},
  journal={IEEE Commun. Mag.}, 
  title={{MIMO} Precoding and Combining Solutions for Millimeter-Wave Systems}, 
  year={Dec. 2014},
  volume={52},
  number={12},
  pages={122-131},
  doi={10.1109/MCOM.2014.6979963}}

@ARTICLE{molisch2017hybrid,
  author={Molisch, Andreas F. and Ratnam, Vishnu V. and Han, Shengqian and Li, Zheda and Nguyen, Sinh Le Hong and Li, Linsheng and Haneda, Katsuyuki},
  journal={IEEE Commun. Mag.}, 
  title={Hybrid Beamforming for Massive {MIMO}: A Survey}, 
  year={2017},
  volume={55},
  number={9},
  pages={134-141},
  doi={10.1109/MCOM.2017.1600400}}

@article{heath2025trihybrid,
      title={The Tri-Hybrid {MIMO} Architecture}, 
      author={Robert W. Heath, Jr. and Joseph Carlson and Nitish Vikas Deshpande and Miguel Rodrigo Castellanos and Mohamed Akrout and Chan-Byoung Chae},
      year={2025},
      journal={arXiv preprint arXiv:2505.21971}
}

@article{liu2025reconfigurable,
      title={Reconfigurable Antenna Arrays: Bridging Electromagnetics and Signal Processing}, 
      author={Mengzhen Liu and Ming Li and Rang Liu and Qian Liu and A. Lee Swindlehurst},
      year={2025},
      journal={arXiv preprint arXiv:2510.17113}
}

@ARTICLE{zhao2026pinching,
  author={Zhao, Jingjing and others},
  journal={IEEE Trans. Commun.}, 
  title={Pinching-Antenna Systems-Enabled Multi-User Communications: Transmission Structures and Beamforming Optimization}, 
  year={2026},
  volume={74},
  number={},
  pages={2316-2330},
  doi={10.1109/TCOMM.2025.3643993}}

@article{Fukuda2022Pinching,
  author  = {Fukuda, A. and Yamamoto, H. and Okazaki, H. and Suzuki, Y. and Kawai, K.},
  title   = {{Pinching antenna: Using a dielectric waveguide as an antenna}},
  journal = {NTT DOCOMO Technical J.},
  volume  = {23},
  number  = {3},
  pages   = {5--12},
  month   = jan,
  year    = {2022}
}

@article{xu2025pinching,
      title={Pinching-Antenna Systems ({PASS}): Power Radiation Model and Optimal Beamforming Design}, 
      author={Xiaoxia Xu and Xidong Mu and Zhaolin Wang and Yuanwei Liu and Arumugam Nallanathan},
      year={2025},
      journal={arXiv preprint arXiv:2505.00218}
}

@ARTICLE{castellanos2026embarcing,
  author={Castellanos, Miguel Rodrigo and Yang, Siyun and Chae, Chan-Byoung and Heath, Robert W.},
  journal={IEEE Trans. Commun.}, 
  title={Embracing Reconfigurable Antennas in the Tri-Hybrid {MIMO} Architecture for {6G} and Beyond}, 
  year={2026, early access, doi:10.1109/TCOMM.2025.3621272},
  volume={74},
  number={},
  pages={381-401},
  doi={10.1109/TCOMM.2025.3621272}}

@ARTICLE{guo2025graph,
  author={Guo, Jia and Liu, Yuanwei and Nallanathan, Arumugam},
  journal={IEEE J. Sel. Areas Commun.}, 
  title={Graph Transformer for Tri-Beamforming in Pinching Antenna Systems ({PASS})}, 
  year={2025, early access, doi:10.1109/JSAC.2025.3645715},
  volume={},
  number={},
  pages={1-1},
  doi={10.1109/JSAC.2025.3645715}}

@ARTICLE{sun2026multiuser,
  author={Sun, Mingjun and Ouyang, Chongjun and Wu, Shaochuan and Liu, Yuanwei},
  journal={IEEE Trans. Wireless Commun.}, 
  title={Multiuser Beamforming for Pinching-Antenna Systems: An Element-Wise Optimization Framework}, 
  year={2026, early access, doi:10.1109/TWC.2025.3625377},
  volume={25},
  number={},
  pages={6538-6552},
  doi={10.1109/TWC.2025.3625377}}

@ARTICLE{ding2025flexible,
  author={Ding, Zhiguo and Schober, Robert and Vincent Poor, H.},
  journal={IEEE Trans. Commun.}, 
  title={Flexible-Antenna Systems: A Pinching-Antenna Perspective}, 
  year={Oct. 2025},
  volume={73},
  number={10},
  pages={9236-9253},
  doi={10.1109/TCOMM.2025.3555866}}

@ARTICLE{shlezinger2021dynamic,
  author={Shlezinger, Nir and Alexandropoulos, George C. and Imani, Mohammadreza F. and Eldar, Yonina C. and Smith, David R.},
  journal={IEEE Wireless Commun.}, 
  title={Dynamic Metasurface Antennas for {6G} Extreme Massive {MIMO} Communications}, 
  year={Apr. 2021},
  volume={28},
  number={2},
  pages={106-113},
  doi={10.1109/MWC.001.2000267}}

@ARTICLE{shi2011an,
  author={Shi, Qingjiang and others},
  journal={IEEE Trans. Signal Process.}, 
  title={An Iteratively Weighted {MMSE} Approach to Distributed Sum-Utility Maximization for a {MIMO} Interfering Broadcast Channel}, 
  year={Sept. 2011},
  volume={59},
  number={9},
  pages={4331-4340},
  doi={10.1109/TSP.2011.2147784}}

@ARTICLE{lin2019hybrid,
  author={Lin, Tian and others},
  journal={IEEE Trans. Commun.}, 
  title={Hybrid Beamforming for Millimeter Wave Systems Using the {MMSE} Criterion}, 
  year={May 2019},
  volume={67},
  number={5},
  pages={3693-3708},
  doi={10.1109/TCOMM.2019.2893632}}

@ARTICLE{bereyhi2025mimo,
      title={{MIMO-PASS}: Uplink and Downlink Transmission via {MIMO} Pinching-Antenna Systems}, 
      author={Ali Bereyhi and Chongjun Ouyang and Saba Asaad and Zhiguo Ding and H. Vincent Poor},
      year={2025},
      journal={arXiv preprint arXiv:2503.03117}
}

@ARTICLE{hou2025on,
  author={Hou, Tianwei and others},
  journal={IEEE Trans. Commun.}, 
  title={On the Performance of Uplink Pinching Antenna Systems ({PASS})}, 
  year={2025, early access, doi:10.1109/TCOMM.2025.3618726},
  doi={10.1109/TCOMM.2025.3618726}}

@ARTICLE{ouyang2025array,
    author={Ouyang, Chongjun and Wang, Zhaolin and Liu, Yuanwei and Ding, Zhiguo},
    journal={IEEE Commun. Lett.}, 
    title={Array Gain for Pinching-Antenna Systems ({PASS})}, 
    year={Jun. 2025},
    volume={29},
    number={6},
    pages={1471-1475},
    doi={10.1109/LCOMM.2025.3566299}}

@ARTICLE{yu2016alternating,
  author={Yu, Xianghao and others},
  journal={IEEE J. Sel. Topics Signal Process.}, 
  title={Alternating Minimization Algorithms for Hybrid Precoding in Millimeter Wave {MIMO} Systems}, 
  year={Apr. 2016},
  volume={10},
  number={3},
  pages={485-500},
  doi={10.1109/JSTSP.2016.2523903}}

@article{zhao2025trihybrid,
      title={Tri-Hybrid Beamforming Design for Fully-Connected Pinching Antenna Systems}, 
      author={Cheng-Jie Zhao and Zhaolin Wang and Hyundong Shin and Yuanwei Liu},
      year={2025},
      journal={arXiv preprint arXiv:2511.14517}
}

@ARTICLE{wiesel2008zero,
  author={Wiesel, Ami and Eldar, Yonina C. and Shamai, Shlomo},
  journal={IEEE Trans. Signal Process.}, 
  title={Zero-Forcing Precoding and Generalized Inverses}, 
  year={Sept. 2008},
  volume={56},
  number={9},
  pages={4409-4418},
  doi={10.1109/TSP.2008.924638}}

@ARTICLE{ngo2013energy,
  author={Ngo, Hien Quoc and Larsson, Erik G. and Marzetta, Thomas L.},
  journal={IEEE Trans. Commun.}, 
  title={Energy and Spectral Efficiency of Very Large Multiuser {MIMO} Systems}, 
  year={Apr. 2013},
  volume={61},
  number={4},
  pages={1436-1449},
  doi={10.1109/TCOMM.2013.020413.110848}}

@ARTICLE{su2022optimal,
  author={Su, Xiaofeng and Jiang, Yi},
  journal={IEEE Wireless Commun. Lett.}, 
  title={Optimal Zero-Forcing Hybrid Downlink Precoding for Sum-Rate Maximization}, 
  year={Mar. 2022},
  volume={11},
  number={3},
  pages={463-467},
  doi={10.1109/LWC.2021.3132338}}

@article{cheng2025performance,
      title={On the Performance of Tri-Hybrid Beamforming Using Pinching Antennas}, 
      author={Zhenqiao Cheng and Chongjun Ouyang and Nicola Marchetti},
      year={2025},
      journal={arXiv preprint arXiv:2511.01099}
}

@article{jiang2025segment,
      title={Segmented Waveguide-Enabled Pinching-Antenna Systems ({SWANs}) for {ISAC}}, 
      author={Hao Jiang and others},
      year={2025},
      journal={arXiv preprint arXiv:2512.07649}
}

@article{ouyang2025uplink,
      title={Uplink and Downlink Communications in Segmented Waveguide-Enabled Pinching-Antenna Systems ({SWANs})}, 
      author={Chongjun Ouyang and others},
      year={Jan. 2026},
      volume={74},
      number={},
      pages={3688-3703},
      doi={10.1109/TCOMM.2026.3653885}}

@ARTICLE{wang2025modeling,
  author={Wang, Zhaolin and others},
  journal={IEEE Trans. Commun.}, 
  title={Modeling and Beamforming Optimization for Pinching-Antenna Systems}, 
  year={Oct. 2025},
  volume={73},
  number={12},
  pages={13904-13919},
  doi={10.1109/TCOMM.2025.3621049}}

@ARTICLE{jiang2025beam,
  author={Jiang, Hao and others},
  journal={IEEE J. Sel. Areas Commun.}, 
  title={Beam Alignment for {MIMO} Fluid Antenna Systems}, 
  year={2025, early access, doi:10.1109/JSAC.2025.3618212},
  doi={10.1109/JSAC.2025.3618212}}
\end{document}